\documentclass[pra,longbibliography,twocolumn,showpacs,nofootinbib,superscriptaddress,notitlepage]{revtex4-1}
\usepackage{amsmath}
\usepackage{amssymb,bm}
\usepackage{color,dsfont}
\usepackage{graphicx} 
\usepackage[colorlinks=true, hyperindex, breaklinks, linkcolor=blue, urlcolor=blue, citecolor=blue]{hyperref} % Physical Review style
\usepackage[normalem]{ulem}
\usepackage{cleveref}
\usepackage{mathrsfs}
\usepackage{subcaption}

\newcommand{\tn}[1]{^{\otimes #1}}
\newcommand{\CPLX}{\mathbb{C}}

\newcommand{\ad}{^\dagger }	% dagger
\newcommand{\bra}[1]{\langle#1|}  % braket
\newcommand{\ket}[1]{|#1\rangle}  % braket
\DeclareMathOperator{\Tr}{\mathrm{Tr}} % trace
\newcommand{\dla}{\ensuremath{\langle\hspace{-.25em}\langle}}
\newcommand{\dra}{\ensuremath{\rangle\hspace{-.25em}\rangle}}
\newcommand{\codepar}[1]{\ensuremath{[\![#1]\!]}}

\begin{document}

\title{Hard decoding algorithm for optimizing thresholds under general Markovian noise}

\author{Christopher Chamberland}
\email{c6chambe@uwaterloo.ca}
\affiliation{
    Institute for Quantum Computing and Department of Physics and Astronomy,
    University of Waterloo,
    Waterloo, Ontario, N2L 3G1, Canada
    }
\author{Joel Wallman}
\email{jwallman@uwaterloo.ca}
\affiliation{
    Institute for Quantum Computing and Department of Physics and Astronomy,
    University of Waterloo,
    Waterloo, Ontario, N2L 3G1, Canada
    }
\author{Stefanie Beale}
\email{sbeale@uwaterloo.ca}
\affiliation{
    Institute for Quantum Computing and Department of Physics and Astronomy,
    University of Waterloo,
    Waterloo, Ontario, N2L 3G1, Canada
    }

\author{Raymond Laflamme}
\affiliation{
    Institute for Quantum Computing and Department of Physics and Astronomy,
    University of Waterloo,
    Waterloo, Ontario, N2L 3G1, Canada
    }
\affiliation{Perimeter Institute, Waterloo, Ontario, N2L~2Y5, Canada}
\affiliation{Canadian Institute For Advanced Research, Toronto, Ontario, M5G~1Z8, Canada}

\begin{abstract}
Quantum error correction is instrumental in protecting quantum systems from 
noise in quantum computing and communication settings. Pauli channels can be 
efficiently simulated and threshold values for Pauli error rates under a 
variety of error-correcting codes have been obtained. However, realistic 
quantum systems can undergo noise processes that differ significantly from 
Pauli noise. In this paper, we present an efficient hard decoding algorithm for 
optimizing thresholds and lowering failure rates of an error-correcting code 
under general completely 
positive and trace-preserving (i.e., Markovian) noise. We use our hard decoding 
algorithm to study the performance of several error-correcting codes under 
various non-Pauli noise models by computing threshold values and failure rates 
for these codes. We compare the performance of our hard decoding algorithm to 
decoders optimized for depolarizing noise and show improvements in thresholds 
and reductions in failure rates by several orders of magnitude. Our hard 
decoding 
algorithm can also be adapted to take advantage of a code's non-Pauli transversal gates to further 
suppress noise. For example, we show that using the transversal gates of the 5-qubit code allows arbitrary rotations around 
certain axes to be perfectly corrected. Furthermore, we show that Pauli 
twirling can increase or decrease the threshold depending upon the code 
properties. Lastly, we show that even if the physical noise model differs 
slightly from the hypothesized noise model used to determine an optimized 
decoder, failure rates can still be reduced by applying our hard decoding 
algorithm.
\end{abstract}


\maketitle

\section{Introduction}

Idealized quantum computers are capable of efficiently factoring very large numbers and 
simulating quantum systems \cite{Shor1994,Zalka1998}. However, 
realistic quantum computers are very sensitive to noise, making
their output unreliable. To overcome the effects of noise, methods for 
error correction and fault-tolerant quantum computation have been developed that allow error rates 
below some threshold value to be arbitrarily suppressed with poly-logarithmic 
overhead \cite{Shor1995,Aharonov1999,Preskill1998,Knill1998}.

In quantum error-correction schemes, ancilla qubits are entangled with 
the set of data qubits that we want to protect. Measuring the ancilla qubits 
produces a (measurement) syndrome that specifies a set of possible errors. 
A recovery operation is then performed in order to correct the error(s) most 
likely to have occurred. A decoding algorithm is an algorithm for determining a good recovery operation for an observed syndrome \cite{Iyer2013}. Note that decoding is elsewhere used to refer to the different process of transferring information from logical to physical qubits.

By the Gottesman-Knill theorem, Pauli channels can be 
efficiently simulated on a classical computer when the underlying quantum 
circuits contain only gates from the Clifford group, qubits prepared in computational basis states and measurements that are performed in the computational basis 
\cite{Aaronson2004}. Simulating 
non-Pauli channels in fault-tolerant architectures is computationally demanding 
and has been done only for small codes 
\cite{Gutierrez2015,Gutierrez2016,Geller2013,Tomita2014}. Assuming perfect 
error correction, that is, perfect preparations of encoded states and syndrome measurements, Rhan \textit{et al.} introduced a technique to obtain the 
effective noise channel after performing error correction \cite{Rahn2002}. The 
technique, based on the process matrix formalism, is applicable to general 
completely positive and trace-preserving (CPTP) noise. Rhan \textit{et al.} 
also showed how to efficiently compute threshold values for concatenated codes 
under a fixed decoder when each qubit is afflicted by CPTP noise. However, the recovery protocols were suboptimal, that is, they did not achieve the best error suppression.

Concatenated codes are formed by encoding each physical qubit of an error correcting code into another code, and the procedure can be repeated recursively. One could obtain the optimal recovery operator by measuring the error syndrome of the full concatenated code. For a code encoding one logical qubit into $n$ physical qubits, the number of syndroms grows as $2^{cn^{l-1}}$ for $l$ levels ($c$ is a constant that depends on the code) making it computationally unfeasible to keep track of all of them \cite{Poulin2006}.

In order to find optimal recovery operators without having to measure the syndromes of the full concatenated code, soft-decoding algorithms were implemented in 
Refs.~\cite{Poulin2006,Fern2008} under the perfect error correction assumptions. In 
Ref.~\cite{Poulin2006}, the entire list of probabilities for all possible 
recoveries conditioned on the observed syndrome were retained and passed on to 
the next level of concatenation in order to implement the optimal recovery operation. 
The method was applied to study thresholds for depolarizing noise. In a message 
passing simulation, the total number of syndromes that must be retained grows 
exponentially with increasing concatenation levels. Therefore, keeping track of 
all syndromes is inefficient.

In Ref.~\cite{Darmawan2016}, again using the perfect error correction 
assumptions, 
Darmawan and Poulin developed a tensor-network algorithm to compute threshold 
values for the surface code under arbitrary local noise. The algorithm allowed 
for the simulation of higher-distance surface codes compared to work done in 
Refs.~\cite{Tomita2014} and resulted in competitive threshold values for the 
studied noise models. However, the algorithm does not use 
non-Pauli transversal gates to its advantage. 

In both the soft-decoding and tensor network approaches, the number of syndromes grows exponentially when increasing the codes distance. Therefore, rather than 
considering all syndrome values, syndromes are sampled from a distribution, 
leading to statistical fluctuations in the reported thresholds. It is possible 
that certain unsampled syndromes could change the behavior of 
the effective noise at the next level in a significant way.

Hard decoding algorithms apply recovery operations independently at each concatenation level based on the measured syndrome (see \cref{subsec:ConcatenatedCodes}). Syndrome information from previous levels are not used to update the recovery maps. This will generally result in a suboptimal recovery protocol. However, hard decoding has the advantage of being constant in the code's distance even when considering \textit{all} syndrome measurements, meaning that the required computational resources to compute the recovery operation remain constant even as the code distance increases exponentially with the number of concatenation levels. 

In this paper we develop a hard decoding algorithm capable of optimizing threshold values and lowering error rates of an error correcting code compared to traditional hard decoding schemes. If the code has non-Pauli single-qubit transversal gates, our algorithm can lead to even further improvements in the computed threshold values and error rates. By single-qubit transversal gates, we refer to gates that can be implemented by applying single-qubit gates to the qubits in a code block. We assume that error correction can be done perfectly so that additional errors are not introduced during the encoding and decoding protocols.

Our hard decoding algorithm is implemented using the process matrix formalism and can be applied to noise models described by general CPTP maps. For noise models which are not depolarizing, the noise behaviour can change between different concatenation levels. Therefore, for a particular syndrome measurement, the best choice of a recovery operator can differ from level to level.

We show that our hard decoding 
algorithm can still lead to reduced error rates even when applied to noise that differs
slightly from the noise used to optimize the recovery maps. This indicates that our 
decoding scheme is robust to perturbative deviations from the assumed noise 
model.

For codes with transversal Clifford gates, we show that applying a Pauli twirl 
to a coherent noise channel results in lower threshold values and higher error rates than those obtained when applying our hard decoding algorithm to the original channel. However, if we only optimize over Pauli recovery maps, the Pauli twirl improves the 
threshold.

The manuscript is structured as follows. We review some preliminary concepts in \cref{sec:ErrorCorrectionProtocolProcessMat}, such as the process matrix formalism (\cref{subsec:ErrorCorrectionProtocol}), stabilizer codes (\cref{subsec:IntroCodes}), logical noise resulting from independent and correlated physical noise (\cref{subsec:ProcessMatUncorrelated} and \cref{subsec:CorrelatedNoiseSec} respectively), logical noise in concatenated codes (\cref{subsec:ConcatenatedCodes}) and threshold hypersurfaces for general noise models (\cref{sec:DefineThreshold}). In \cref{sec:OptimizedDecoding} we describe our hard decoding algorithm for optimizing threshold values of Markovian noise models. In \cref{sec:NumericalThresholdCal} we describe how to numerically calculate threshold hypersurfaces for both symmetric decoders and decoders obtained from our hard decoding algorithm. 

We then present the results of numerical simulations of the 5-qubit code, Steane's 7-qubit code, Shor's 
9-qubit code and the surface-17 code. For each code (excluding the surface-17 
code), thresholds and infidelities using our hard decoding optimization 
protocol are computed for amplitude-phase damping noise (\cref{sec:AmpPhaseDamp}) and coherent 
noise (\cref{sec:CoherentErrorThresholdAnalysis}). The concept of infidelity is defined later in the manuscript in \cref{eq:infidelityConcept}. For the same noise models, we consider level-1 infidelities of the 
surface-17 code. We also consider thresholds and infidelities for the 7-qubit 
code where the noise model was described by two-qubit correlated dephasing 
noise (\cref{sec:CorrelatedNoiseThresholdAnalysis}). In \cref{subsec:PauliTwirl}, we compute thresholds of the Steane code for a coherent error noise channel and its Pauli twirled counterpart (using both logical Clifford corrections and Pauli only corrections). Lastly, we study the robustness of our decoding algorithm to small unknown perturbations of a noise channel (\cref{subsec:SensitivityDecoder}).

The amplitude-phase damping threshold and infidelity plots can be found 
in \cref{fig:AmpDampPlot}. Applying our optimized hard decoding algorithm can 
more than double thresholds and reduce infidelities by more than two orders 
of magnitude.

For coherent error noise, threshold plots are given 
in \cref{fig:CoherentPlot} and infidelity plots are given in 
\cref{fig:CoherentInfidelities}.  For certain rotation axes, our optimized 
hard decoding algorithm results in errors that are correctable for all rotation 
angles. In some regimes, infidelities can be reduced by several orders of 
magnitude. For all the aforementioned noise models, the 
5-qubit code consistently achieves higher thresholds and lower error rates 
compared to the 7 and 9-qubit codes. There is one exception where the Hadamard 
transform of the 9-qubit code outperforms the 5-qubit code for amplitude-phase 
damping noise in a small regime. For most studied noise models, the 7-qubit 
code achieves higher threshold values and lower error rates than the 9-qubit 
code, with the exception of rotations near the $y$-axis due to the asymmetries in the Shor codes stabilizer generators.

The threshold plot comparing a coherent error noise channel to its Pauli twirled counterpart is shown in \cref{fig:7qubitTwirlPlots}. By performing Clifford corrections, the coherent noise channel outperforms its Pauli twirled counterpart for all sampled rotation axes. Lastly, plots showing the robustness of our decoding algorithm to small unknown perturbations are shown in \cref{fig:7qubitPerturbedInfidelityPlots}. In certain regimes, applying our decoding algorithm results in lower logical failures rates even if the noise model is not perfectly known.

\section{Stabilizer codes and the process matrix formalism}
\label{sec:ErrorCorrectionProtocolProcessMat}

We begin by outlining the formalism we use to simulate the performance of 
concatenated codes under general CPTP noise. We review the process matrix 
formalism for CPTP maps in \cref{subsec:ErrorCorrectionProtocol} and general 
stabilizer codes (an important class of quantum error-correcting codes) in 
\cref{subsec:IntroCodes}. We derive expressions for the process matrix 
conditioned on observing a specific measurement syndrome for independent 
single-qubit noise in \cref{subsec:ProcessMatUncorrelated}, and for two-qubit 
correlated noise in \cref{subsec:CorrelatedNoiseSec}. We then define 
thresholds for a noise model in \cref{sec:DefineThreshold} and 
define some fixed decoders in \cref{sec:SymmetricDecoding}.

For clarity, we always use Roman font for operators acting on 
$\mathbb{C}^d$ (e.g., a unitary $U$), calligraphic font for a channel acting on 
the operator space (i.e., a superoperator, e.g., $\mathcal{U}(\rho) = U\rho 
U\ad$) and bold calligraphic font for the matrix representation of a channel 
(e.g., $\boldsymbol{\mathcal{U}}$).

\subsection{Process matrix formalism for noise at the physical level}
\label{subsec:ErrorCorrectionProtocol}

A CPTP noise channel $\mathcal{N}$ acting on a state $\rho$ can be written 
in terms 
of its Kraus operator decomposition
\begin{align}
\mathcal{N}(\rho)=\sum_{j} A_{j}\rho A_{j}^{\dagger},
\label{eq:KrausOp}
\end{align}
where $\sum_{j} A_{j}^{\dagger}A_{j}=I$ for trace-preserving channels 
\cite{Kraus1983,Nielsen_Chuang}. 
Alternatively, \cref{eq:KrausOp} can be rewritten as a matrix product using the 
process matrix formalism. To do so, note that any matrix $M\in\CPLX^{d \times d}$ can be 
expanded as
\begin{align}
M = \sum_i M_i B_i,
\label{eq:DensityExpand}
\end{align}
where $\boldsymbol{B} =\{B_i\}$ is a trace-orthonormal basis for the space of 
density matrices, that is, $\Tr (B_i\ad B_j) = \delta_{i,j}$, and $M_i = 
\Tr 	
(B_i\ad M)$. We exclusively study multi-qubit channels and so set the $B_i$ to be 
the normalized Pauli matrices, $\boldsymbol{B} = \boldsymbol{\sigma} = 
(I,X,Y,Z)/\sqrt{2}$ for a single qubit, and $\boldsymbol{B} = 
\boldsymbol{\sigma}\tn{n}$ for $n$ qubits.

We then define a map $|.\dra:\CPLX^{d\times d}\to\CPLX^{d^2}$ by setting 
$|B_j\dra = \textbf{e}_j$, where $\{\textbf{e}_j\}$ is the canonical unit basis 
of $\CPLX^{d^2}$, and extend the map linearly so that
\begin{align}
|M\dra = \sum_j M_j |B_j\dra = \left( 
\begin{array}{c}                                                  
\Tr[B_1\ad M] \\
\vdots \\
\Tr[B_{d^2}\ad M]
\end{array} \right).
\label{eq:VecRep}
\end{align} 
Defining $\dla M| = |M\dra\ad$, we have $\dla M|N\dra = \Tr (M\ad N)$.

Because quantum channels are linear,
\begin{align}
|\mathcal{N}(\rho)\dra&= \sum_j \rho_j |\mathcal{N}(B_j)\dra \notag\\
&= \sum_{i,j} \rho_i |\mathcal{N}(B_j)\dra\!\dla B_j|B_i\dra \notag\\
&= \Bigl(\sum_j |\mathcal{N}(B_j)\dra\!\dla B_j|\Bigr) \Bigl(\sum_{i}\rho_{i} |B_{i}\dra \Bigr) \notag\\
&= \boldsymbol{\mathcal{N}}|\rho\dra
\label{eq:Nvec}
\end{align}  
where $\rho_{j}=\dla B_j|\rho\dra$ are the expansion coefficients of $\rho$ and we used $\dla B_j|B_i\dra = \Tr (B_j\ad B_i) = 
\delta_{j,i}$. We implicitly defined the matrix representation
$\boldsymbol{\mathcal{N}}$ of the 
quantum channel $\mathcal{N}$.

\subsection{Stabilizer codes}
\label{subsec:IntroCodes}

We now review stabilizer codes~\cite{Gottesman2010}. 
An \codepar{n,k,d} stabilizer code $C$ corresponds to the unique 
subspace $\mathcal{H}_C$ of the $n$-qubit Hilbert space $\mathcal{H}$ which is 
the $+1$ eigenspace of an Abelian subgroup
$\mathcal{S}$ ($-I \notin \mathcal{S}$) of the $n$-qubit Pauli group. The stabilizer group 
$\mathcal{S}$ 
is generated by a set of $n-k$ mutually-commuting $n$-qubit Pauli operators 
$\{g_{1},g_{2},\hdots ,g_{n-k}\}$. Non-identity Pauli operators which commute with all elements of $\mathcal{S}$ act non-trivially (i.e. differs from the identity) on at least $d$ qubits. Defining $N(\mathcal{S})$ as the set of all Pauli operators that commute with $\mathcal{S}$, any Pauli in $N(\mathcal{S}) \setminus \mathcal{S}$ acts as a logical Pauli operator on encoded states. 
For this paper, we only consider the stabilizer codes in 
\cref{tab:StabilizerGeneratorsLists}, and so set $k=1$ for the remainder of 
this section.

We assume that states in $\mathcal{H}_2$ (the Hilbert space of unencoded 
states) can be encoded in $\mathcal{H}_C$ by an encoding map 
$\mathcal{E}$ and decoded back to $\mathcal{H}_2$ by  
the adjoint map $\mathcal{E}^{\dagger}$. We consider the case where the encoding and decoding protocols can be done perfectly without introducing additional errors, so that $\ket{\psi} = \alpha \ket{0} + \beta 
\ket{1}$ is encoded to $\ket{\overline{\psi}} = \alpha \ket{\overline{0}} + \beta 
\ket{\overline{1}} \in \mathcal{H}_C$ and
\begin{align}
\mathcal{E}(\rho_{in}) = B \rho_{in} B^{\dagger}
\label{eq:EncodingMap}
\end{align}
with $B = \ket{\overline{0}}\bra{0} + \ket{\overline{1}}\bra{1}$ and some 
abuse of notation. Since $(1/|\mathcal{S}|)\sum_{k}\mathcal{S}_{k}$ acts as the projector onto the codespace ($|\mathcal{S}|$ is the total number of elements in the stabilizer group) and representing $\overline{\tau}$ as the logical version of $\tau$ ( $\tau \in \boldsymbol{\sigma}$), we define
\begin{align}
E_{\tau} = \mathcal{E}(\tau) = 
\frac{1}{|\mathcal{S}|}\sum_{S\in\mathcal{S}} S\overline{\tau},
\label{eq:EncodingDef}
\end{align} 
so that $E_{\tau}$ implements $\tau$ in $\mathcal{H}_{C}$ and vanishes elsewhere. 

We also assume that syndrome measurements and recovery maps $\mathcal{R}$ are 
perfect, so that the only errors are due to a noise process $\mathcal{N}$ 
acting on $\mathcal{H}$. More details about fault-tolerant encoding 
and measurements can be found in, for example, 
Refs.~\cite{Gottesman2010,Aliferis2008,Paetznick2011,Chamberland2016, 
Chamberland2016a}. 

Suppose a physical Pauli error $E$ occurs on a system in the encoded state 
$\ket{\overline{\psi}}\in\mathcal{H}_C$. Measuring the stabilizer generators 
yields the syndrome $l=l_{1}l_{2}\hdots l_{n-k}$ where 
\begin{align}
l_i= \begin{cases}
0 & \mbox{if } [E,g_i]=0\\
1 & \mbox{if } \{E,g_i\}=0,
\end{cases}
\end{align}
$[A,B] = AB-BA$ and $\{A,B\}= AB+BA$.
Let $Q_l$ be the set of physical Pauli errors that give the syndrome $l$, which 
are all of size $\lvert Q_l \rvert = 2^{2n}/2^{n-k} = 2^{n+k}$. When the 
syndrome $l$ is measured, a recovery operator $R_l\in Q_l$ is chosen and 
applied to the state $E\ket{\overline{\psi}}$, returning it to the code space. 
If $R_l E \in \mathcal{S}$, then the correct state is recovered and the error is removed. 
Otherwise, $R_l E\ket{\overline{\psi}}$ will differ from 
$\ket{\overline{\psi}}$ by a logical Pauli operator~\cite{Gottesman2010}. The 
desired outcome of decoding is to find a set of recovery operators which result in the 
highest probability of recovering the original input state under a given noise 
model.

As an example, the stabilizer generators for the 3-qubit code protecting 
against bit-flip errors are $S_{1} = Z_{1}Z_{2}$ and $S_{2} = Z_{2}Z_{3}$. It 
can be verified that the errors $X_{1}$ and $X_{2}X_{3}$ produce the syndrome 
$l=10$ so that $Q_{10} = \{X_{1}, X_{2}X_{3} \}$. Therefore, if the measured 
syndrome is $l=10$, one can either choose $X_{1}$ or $X_{2}X_{3}$ to implement 
the recovery. The particular choice can influence the fidelity of the encoded 
qubit. For instance, for uncorrelated noise models where single-weight errors 
are more likely, the better choice for the recovery operator would be $R_{10} = 
X_{1}$.

\begin{table}[t]
\begin{tabular}{ c|c|c|c}
 5-qubit code & Steane code & Z-Shor code & Surface-17 code \\ \hline
     XZZXI      &   IIIZZZZ     & ZZIIIIIII & ZIIZIIIII \\
     IXZZX      &   IZZIIZZ     & ZIZIIIIII & IZZIZZIII \\
     XIXZZ      &   ZIZIZIZ     & IIIZZIIII & IIIZZIZZI \\
     ZXIXZ      &    IIIXXXX   & IIIZIZIII & IIIIIZIIZ \\
				&   IXXIIXX     & IIIIIIZZI & XXIXXIIII \\
                &   XIXIXIX    & IIIIIIZIZ & IXXIIIIII \\
                &                 &\ XXXXXXIII \ &  IIIIXXIXX \\
                &                 &\ IIIXXXXXX \ &  IIIIIIXXI  \\\hline
$\langle C_{\pi/3},X,Z\rangle$ & $\langle H,S\rangle$ & $\langle 
X,Z\rangle$ & $\langle X,Z\rangle$
\end{tabular}
\caption{Stabilizer generators (top) and the group $\mathscr{L}$ of 
single-qubit transversal logical operations (bottom) for the 5-qubit 
code~\cite{Laflamme1996}, Steane's 7-qubit code~\cite{Steane1996}, Shor's 
9-qubit 
code~\cite{Shor1995}, and the surface-17 code~\cite{Tomita2014}, where $H$ and 
$S$ are 
the Hadamard and phase gates respectively, 
$C_{\pi/3} = \exp[i \pi(X+Y+Z)/3\sqrt{3}]\propto SH$, and $\langle.\rangle$ denotes the group 
generated by the argument. We also consider the $X$-Shor code, obtained from the 
$Z$-Shor code by mapping $X\leftrightarrow Z$. For each code, the logical 
operators are $X_{L} = X^{\otimes n}$ and $Z_{L} = Z^{\otimes n}$. We only 
consider the surface-17 code at the first level, as surface codes are not scaled 
up by concatenation. The surface-17 code is so named as it consists of 9 data qubits 
and 8 ancilla measurement qubits, and is equivalent to the other 2-D 
configuration with 9 data qubits in Ref.~\cite{Tomita2014} under the assumption 
of 
perfect measurements.}
\label{tab:StabilizerGeneratorsLists}
\end{table}

\subsection{Effective process matrix at the logical level}
\label{subsec:ProcessMatUncorrelated}

The process of encoding, applying the physical noise $\mathcal{N}$ to the 
encoded state, implementing the appropriate recovery maps for the measured syndrome $l$ and decoding yields the effective 
single-qubit channel 
\begin{align}
\mathcal{G}(\mathcal{N},R_{l}) =\mathcal{E}^{\dagger}\circ\mathcal{R}_{l} \circ 
\mathcal{N} \circ \mathcal{E},
\end{align}
where $\mathcal{R}_l$ includes the measurement 
update and the recovery map $R_{l} \in Q_{l}$. We now outline how this effective channel can be 
represented in the process matrix formalism, mostly following 
Ref.~\cite{Rahn2002} 
with a straightforward generalization to consider individual syndromes. The 
states before encoding and after decoding, $\rho_\mathrm{in}$ and 
$\rho_\mathrm{out}$ respectively, are related by
\begin{align}
| \rho_\mathrm{out} \dra = \boldsymbol{\mathcal{G}}(\mathcal{N},R_{l}) 
|\rho_\mathrm{in} \dra,
\label{eq:InitialFinal}
\end{align}
where the process matrix representation of $\boldsymbol{\mathcal{G}}(R_{l})$ is
\begin{align}
\boldsymbol{\mathcal{G}}(\mathcal{N},R_{l}) = 
\sum_{\sigma\in\boldsymbol{\sigma}} 
|\mathcal{G}(\mathcal{N},R_{l}) (\sigma)\dra\!\dla \sigma|
\label{eq:MatRepG}
\end{align}
by \cref{eq:Nvec}. The entries of the process matrix are
\begin{align}
\boldsymbol{\mathcal{G}}_{\sigma\tau}(\mathcal{N},R_{l})
&= \dla \sigma | \mathcal{G}(\mathcal{N},R_{l})(\tau)\dra \notag\\
&= \dla \mathcal{E}(\sigma) |\mathcal{R}_l \circ \mathcal{N}(E_\tau)\dra 
\notag\\
&= \dla E_\sigma |\mathcal{R}_l \circ \mathcal{N}(E_\tau)\dra.
\label{eq:Gvec}
\end{align}

By the Born rule, the probability of the syndrome $l$ occurring is $p(l) = 
\Tr(\mathscr{P}_{l}\mathcal{N}(\rho_{\rm in}))$ where the projection operator 
for the syndrome $l$ is
\begin{align}
\mathscr{P}_l = \prod_{j=1}^{n-k}\frac{1}{2}(I+(-1)^l g_j).
\label{eq:ProjectorEquation}
\end{align}
Note that from \cref{eq:EncodingDef}, $E_{\tau} = \mathscr{P}_{0} 
\overline{\tau}$. With the corresponding recovery operator $R_{l}$, the 
transformation on the process matrix can be obtained by implementing the von 
Neumann-L{\"u}ders update rule \cite{Baumann2016} resulting in

\begin{align}
\boldsymbol{\mathcal{G}}_{\sigma\tau}(\mathcal{N},R_l) 
&= \frac{1}{p(l)} \dla E_{\sigma} | R_{l} \mathscr{P}_{l} \mathcal{N}(E_{\tau}) \mathscr{P}_{l}^{\dagger} R_{l}^{\dagger} \dra \notag\\
&= \frac{1}{p(l)} \dla \mathscr{P}_{l}^{\dagger} R_{l}^{\dagger} E_{\sigma} R_{l} \mathscr{P}_{l} | \mathcal{N}(E_{\tau}) \dra
\label{eq:Gmapsigma}
\end{align}

Following Ref.~\cite{Rahn2002}, \cref{eq:Gmapsigma} can be further simplified 
by noting that $R_{l}^{\dagger} E_{\sigma}R_{l}$ is a map from the space 
projected by $\mathscr{P}_{l}$ to itself and vanishes elsewhere so that 
$\mathscr{P}_{l}^{\dagger}R_{l}^{\dagger} E_{\sigma} 
R_{l}\mathscr{P}_{l}=R_{l}^{\dagger} E_{\sigma} R_{l}$. Defining
\begin{align}
D_{\sigma}^{(l)} \equiv R_{l}^{\dagger} E_{\sigma} R_{l},
\label{eq:Dsigma}
\end{align}
we arrive at 
\begin{align}
\boldsymbol{\mathcal{G}}_{\sigma \tau}(\mathcal{N},R_{l}) =\frac{1}{p(l)} 
\dla D_{\sigma}^{(l)} | \mathcal{N}(E_{\tau}) \dra.
\label{eq:GcomponentFinal}
\end{align}

In the remainder of this section we will assume that the noise is uncorrelated, so that it takes the form $\mathcal{N} = \mathcal{N}^{(1)}\otimes \ldots\otimes\mathcal{N}^{(n)}$ where $\mathcal{N}^{(i)}$ is the process matrix for the physical noise acting on qubit $i$. As in \cref{eq:DensityExpand}, we can expand $E_{\tau}$ and 
$D_{\sigma}^{(l)}$ as
\begin{align}
E_\tau &= \sum_{\mu_{i} \in \tilde{\boldsymbol{B}}} \alpha_{\{ \mu_{i} 
\}}^\tau \mu_{1} \otimes \hdots \otimes \mu_{n},
\label{eq:EsigmaTensorProduct}
\end{align}
\begin{align}
D_{\sigma}^{(l)} &= \sum_{\nu_{i} \in \tilde{\boldsymbol{B}}} \beta_{\{ \nu_{i} 
\}}^{\sigma}(R_l) \nu_{1} \otimes \hdots \otimes \nu_{n},
\label{eq:EDsigmaTensorProduct}
\end{align}
where $\tilde{\boldsymbol{B}} \subset \boldsymbol{B}$ only has support over products of the stabilizer 
group and logical operators from \cref{eq:EncodingDef}. For an 
operator of the form $U = \pm \mu_{1} \otimes \hdots \otimes \mu_{n}$ and using 
the notation of Ref.~\cite{Rahn2002}, we define the function $\phi(U) = \mu_{1} 
\otimes 
\hdots \otimes \mu_{n}$ and $a(U) \in \{0, 1\}$ such that $U = 
(-1)^{a(U)}\phi(U)$. Substituting \cref{eq:EncodingDef} into
\cref{eq:EsigmaTensorProduct} gives
\begin{align}
\alpha_{\phi(S\overline{\tau})}^{\tau} = \frac{1}{2^{\frac{n}{2}-1}} 
(-1)^{a(S \overline{\tau})}.
\label{eq:AlphaCoefficient}
\end{align}

The $\alpha$ coefficient takes into account the 
overall sign of the product between elements in the stabilizer group and logical operators (for 
example, the code with $XX$ and $ZZ$ stabilizers also has $(XX)(ZZ) = 
-YY$ as a stabilizer). The factor of $\frac{1}{2^{\frac{n}{2}-1}}$ comes from choosing a trace-orthonormal basis.

The $\beta$ coefficient can be obtained by substituting \cref{eq:EncodingDef} into \cref{eq:Dsigma}, commuting $R_{l}$ to the left, using $R_{l}^{\dagger} R_{l} = I$ and setting the result equal to \cref{eq:EDsigmaTensorProduct}. Defining $\eta(A,B)= \pm 1$ for $AB = \pm BA$, we obtain
\begin{align}
\beta_{\phi(S_{k}\overline{\sigma})}^{\sigma}(R_{l})=\alpha_{\phi(S_{k}\overline{\sigma})}^{\sigma}
 \eta(R_{l},S_{k})\eta(R_{l},\overline{\sigma}).
\label{eq:BetaIndividualSyn}
\end{align}
Therefore, for a particular error syndrome $l$, picking different recovery operators from the set $Q_{l}^{\dagger}$ will, in general, yield different values for the 
coefficient $\beta$. This will, in turn, result in different effective noise dynamics. Closed form expressions for $\alpha$ and $\beta$ are given in \cref{app:AalphaBeta}.

Substituting \cref{eq:EDsigmaTensorProduct} into \cref{eq:GcomponentFinal}, we obtain

\begin{align}
\boldsymbol{\mathcal{G}}_{\sigma \tau}(\mathcal{N},R_{l}) = \frac{1}{p(l)}
\sum_{\{\mu_{i}\},\{\nu_{i}\}} \beta_{\{\nu_{i}\}}^{\sigma}(R_l) 
\alpha_{\{\mu_{i}\}}^{\tau}\prod_{i=1}^{n}\boldsymbol{\mathcal{N}}_{\nu_{i} 
\mu_{i}}^{(i)},
\label{eq:GmatrixIndividualSyndromes}
\end{align}
where the sum is over all elements in the stabilizer group and 
$\boldsymbol{\mathcal{N}}_{\nu_{i} \mu_{i}}^{(i)} = \Tr[\nu_{i} \mathcal{N}^{(i)}(\mu_{i})]$. 

The effective noise channel can be obtained by averaging \cref{eq:GmatrixIndividualSyndromes} over all syndrome measurements. Defining $\beta^{\sigma}_{\{\nu_{i} \}} \equiv \sum_{l}\beta^{\sigma}_{\{\nu_{i} \}}(R_{l})$, we have 

\begin{align}
\boldsymbol{\mathcal{G}}_{\sigma \tau}(\mathcal{N})
&= \sum_{l} p(l) \boldsymbol{\mathcal{G}}_{\sigma \tau}(\mathcal{N},R_{l}) 
\notag\\
& = \sum_{\{\mu_{i}\},\{\nu_{i}\}} \beta_{\{\nu_{i}\}}^{\sigma} \alpha_{\{\mu_{i}\}}^{\tau}\prod_{i=1}^{n}\boldsymbol{\mathcal{N}}_{\nu_{i} 
\mu_{i}}^{(i)},
\label{eq:Gmatrix}
\end{align} 
and we will refer to ${\mathcal{G}}$ as the effective process matrix for the noise channel $\mathcal{N}$. Note that the normalization factor $1/p(l)$ that appears when implementing the von Neumann-L{\"u}ders update rule gets cancelled when averaging over all syndrome measurements. For simplicity, and in the remaining sections of this paper, when referring to process matrices for individual syndrome measurements as in \cref{eq:GmatrixIndividualSyndromes}, we will omit the normalization factor. 

When considering concatenated codes in \cref{subsec:ConcatenatedCodes}, it will prove useful to define the coding map $\Omega^{C}$ for a code $C$ as 

\begin{align}
\Omega^{C} : \mathcal{N} \to \mathcal{G}(\mathcal{N})= \mathcal{E\ad} \circ 
\mathcal{R} \circ \mathcal{N} \circ \mathcal{E},
\label{eq:CodingMap}
\end{align} 
where the matrix representation of $\mathcal{G}$ is obtained from \cref{eq:Gmatrix}. Note that in \cref{eq:CodingMap}, $\mathcal{R}$ includes the measurement update and recovery map averaged over all syndrome measurements. The coding map relates the effective noise dynamics at the logical level resulting from the error correction protocol to the noise dynamics occurring at the physical level. 

\subsection{Process matrix for two-qubit correlated noise}
\label{subsec:CorrelatedNoiseSec}
In this paper we also consider noise models 
where nearest-neighbor two-qubit correlations occur. More specifically, 
we will consider a noise channel of the form 
\begin{align}
&\mathcal{N}(\rho^{\otimes n}) = \mathcal{N}^{(1)}(\rho)^{\otimes n}+ \nonumber \\
&p_{2}(\sum_{j=1}^{n-1}\mathcal{N}_{j,j+1}^{(2)}(\rho^{\otimes n})+\mathcal{N}_{1,n}^{(2)}(\rho^{\otimes n})),
\label{eq:CorrelatedNoise}
\end{align} 
where $\mathcal{N}^{(1)}$ corresponds to local uncorrelated noise and with probability $p_{2}$, $\mathcal{N}_{j,j+1}^{(2)}(\rho^{\otimes n})=Z_{j}Z_{j+1}\rho^{\otimes n}Z_{j+1}Z_{j}$ applies phase-flip operators to qubits $j$ and $j+1$.  
For noise models of this form, the process matrix describing the effective noise is given by 
\begin{align}
&\boldsymbol{\mathcal{G}}_{\sigma \tau}(\mathcal{N}) = 
\boldsymbol{\mathcal{G}}_{\sigma 
\tau}([\mathcal{N}^{(1)}]^{\otimes n})+ 
\nonumber \\
& p_{2}\sum_{\{\mu_{i}\},\{\nu_{i}\}}\beta_{\{\nu_{i}\}}^{\sigma}\alpha_{\{\mu_{i}\}}^{\tau}(\sum_{j=1}^{n-1}\prod_{\substack{i=1 \\ i\notin \{j,j+1\}}}^{n}\boldsymbol{\mathcal{Z}}^{(2)}_{\nu_{j} \mu_{j}}\boldsymbol{\mathcal{Z}}^{(2)}_{\nu_{j+1} \mu_{j+1}}\mathcal{I}_{\nu_{i}\mu_{i}}+ \nonumber \\
&\prod_{i=2}^{n-1}\boldsymbol{\mathcal{Z}}^{(2)}_{\nu_{1} \mu_{1}}\boldsymbol{\mathcal{Z}}^{(2)}_{\nu_{n} \mu_{n}}\mathcal{I}_{\nu_{i}\mu_{i}}),
\label{eq:CorrelatedNoiseProcessMat}
\end{align} 
where $\mathcal{Z}(\rho) = Z\rho Z$ in keeping with our standard notation for channels.
The contribution from correlated noise appears in the second term of 
\cref{eq:CorrelatedNoiseProcessMat}. 

\subsection{Effective noise channels for concatenated codes} 
\label{subsec:ConcatenatedCodes}
Concatenation is the process of encoding each of the $n$ physical qubits encoded in an inner code $C_{1}$ into an outer code $C_{2}$. One can go to arbitrary levels of concatenation by recursively applying this procedure.

More formally, we consider an $m$-qubit code $C^{out}$ with encoding map $ \mathcal{E}^{out}$ which will form the 
outer code, and an $n$-qubit code $C^{in}$ with encoding map $ \mathcal{E}^{in}$ which will form the inner code. The logical qubit $\rho_{0}$ is first encoded using $C^{out}$, and 
afterwards each of the $m$ qubits are encoded using the 
code $C^{in}$. The composite encoding map is given by 
\begin{align}
\tilde{\mathcal{E}}=(\mathcal{E}^{in})^{\otimes m}\circ \mathcal{E}^{out}.
\label{eq:CompositeEncoding}
\end{align} 
Throughout this paper we will implement a hard decoding scheme, which applies a 
recovery operation independently at each concatenation level 
\cite{Gottesman2010}. Each code block is thereby corrected 
based on the inner code. The entire register is then corrected based on the 
outer code. We denote the $mn$-qubit code with the effective 
encoding map $\tilde{\mathcal{E}}$ by $C^{out}(C^{in})$. The 
procedure for choosing a decoding map for a given noise model described by a 
CPTP map will be addressed in 
\cref{sec:OptimizedDecoding}. 

Let $\mathcal{G}$ describe the effective dynamics of $C^{in}$ where the physical noise dynamics are described by $\mathcal{N}$. To obtain the effective noise dynamics of $\tilde{\mathcal{G}}$ for the code $C^{out}(C^{in})$, we assume that all $n$-qubit blocks evolve according to $\mathcal{N}$ so that the $mn$-qubit code evolves according to
\begin{align}
\tilde{\mathcal{N}}=\mathcal{N}^{\otimes m}.
\label{eq:EvolveMN}
\end{align} 
For convenience, we define $\mathcal{E}\ad_{\mathcal{R}} \equiv \mathcal{E}\ad \circ 
\mathcal{R}$ so that $\mathcal{E}\ad_{\mathcal{R}}$ includes both the recovery and 
decoding step. In Ref.~\cite{Rahn2002}, it was shown that with the above 
assumptions $\tilde{\mathcal{G}}$ is given by
\begin{align}
\tilde{\mathcal{G}}=(\mathcal{E}\ad)_{\mathcal{R}}^{out}\circ \mathcal{G}^{\otimes m}\circ \mathcal{E}^{out}.
\label{eq:Gtilde}
\end{align} 
From \cref{eq:CodingMap}, the above equation can be written as 
\begin{align}
\tilde{\mathcal{G}}=\Omega^{C^{out}}(\mathcal{G})=\Omega^{C^{out}}(\Omega^{C^{in}}(\mathcal{N})).
\label{eq:GtildeMapComp}
\end{align} 
For uncorrelated noise, we conclude that the effective channel for the code $C^{out}(C^{in})$ can be computed in the same way that lead to \cref{eq:Gmatrix} by replacing $\mathcal{N}$ with $\mathcal{G}$ for the code $C^{out}$. The concatenated code $C^{out}(C^{in})$ can then be described by the composition of maps
\begin{align}
\Omega^{C^{out}(C^{in})}=\Omega^{C^{out}}\circ \Omega^{C^{in}}.
\label{eq:CompMap}
\end{align}
The above equation can be easily generalized to the concatenation of codes in the set $\{C_{1},C_{2},\hdots ,C_{n}\}$ yielding the map
\begin{align}
\Omega^{C_{1}(C_{2}(\hdots C_{n}))}=\Omega^{C_{1}}\circ \Omega^{C_{2}}\circ \hdots \circ \Omega^{C_{n}}.
\label{eq:CompMapGeneral}
\end{align}
For the particular case where the same code $C_{i}=C$ ($i\in \{1,2,\hdots 
,t\}$) is used at $t$ levels of concatenation, we define
\begin{align}
\mathcal{G}^{(t)}(\mathcal{N})=\Omega^{C_{1}(C_{2}(\hdots 
C_{t}))}(\mathcal{N}).
\label{eq:GdefConcatT}
\end{align} 

For correlated noise as in \cref{subsec:CorrelatedNoiseSec}, we 
cannot in general write the map for the code $C^{out}(C^{in})$ 
as a composition of maps for the code $C^{out}$ and $C^{in}$. However, in 
this paper we will assume that when the code is concatenated, no correlations 
occur between different code blocks. Only qubits within each code block undergo 
correlated noise described by \cref{eq:CorrelatedNoise}. The noise 
dynamics for each code block of the code $C^{out}$ will thus be described by 
the effective noise dynamics of \cref{eq:CorrelatedNoiseProcessMat} and the 
analysis leading to \cref{eq:CompMapGeneral} will also apply in this case.
This situation could be realized if the physical qubits in each 
lowest-level code are contained in individual nodes of a distributed quantum 
computer and is a good approximation if correlations decay exponentially with 
the separation between physical qubits.

\subsection{Thresholds for noise models}
\label{sec:DefineThreshold}

A fixed noise process $\mathcal{N}$ is correctable by a concatenated code 
$C$ if successive levels of concatenation eventually remove the error 
completely for arbitrary input states, that is, if
\begin{align}
\lim_{t\to \infty} \boldsymbol{\mathcal{G}}^{(t)}(\mathcal{N})= I_4,
\label{eq:Gthreshold}
\end{align}
where $\boldsymbol{\mathcal{G}}^{(t)}(\mathcal{N})$ is as defined in 
\cref{eq:GdefConcatT} and $I_{4}$ is the $4 \times 4$ identity matrix. 
(Formally, we could also require the error rate to decrease 
doubly-exponentially when quantified by an appropriate 
metric~\cite{Nielsen_Chuang}, 
however, we do not verify this requirement.)

A threshold for a code is defined relative to an $m$-parameter noise model, 
that is, a family $\mathscr{N}=\{\mathcal{N}_p:p\in[0,1]^m\}$ of noise 
processes such that $\mathcal{N}_0 = \mathcal{I}$. The $\mathscr{N}$-threshold 
for a code $C$ is the hypersurface of the largest volume in $[0,1]^m$ 
containing only correctable noise processes and the origin, with the faces of 
$[0,1]^m$ removed.

\begin{figure}
\centering
\includegraphics[width=0.5\textwidth]{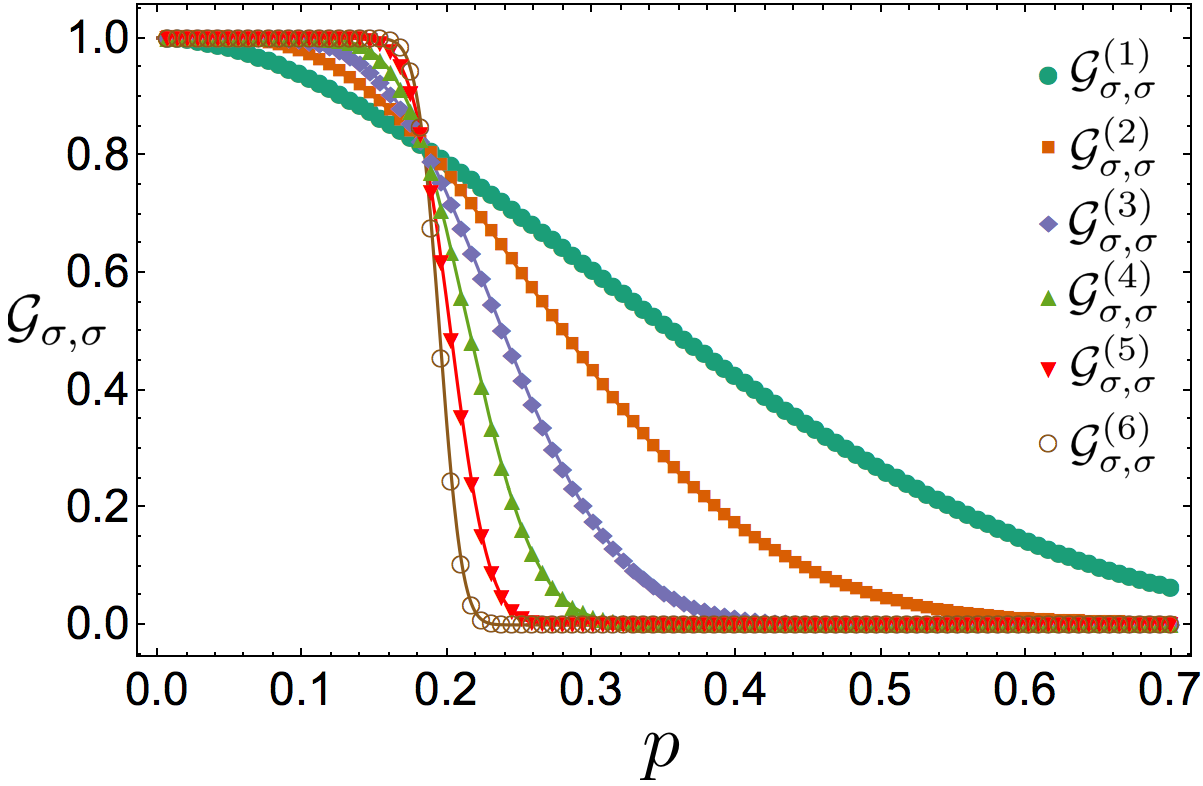}
\caption{Representative plot of the smallest diagonal component $\boldsymbol{\mathcal{G}}_{\sigma,\sigma}$ of the process matrix 
for a noise model parametrized by $p$. As functions of $p$, the diagonal components of the process matrix approach a 
step-function as the number of concatenation levels approaches infinity. The 
threshold is the smallest value $p_{th}$ such that 
$\lim_{t\to\infty}\boldsymbol{\mathcal{G}}^{(t)}(\mathcal{N}_p)= 
I_4$ for all $p\leq p_{th}$.}
\label{fig:ProcessMatDiagFig}
\end{figure}

The typical behavior of the diagonal components of the process matrix for a 
1-parameter noise model is illustrated in \cref{fig:ProcessMatDiagFig}. The 
diagonal components converge to one (zero) below (above) threshold, while the 
off-diagonal components converge to zero.

\subsection{Specific decoders}\label{sec:SymmetricDecoding}

The effective noise acting on a logical qubit is highly dependent upon both the 
physical noise processes and the choice of recovery operators for each syndrome 
(cf. \cref{eq:Gmatrix}).

One decoder that will be very useful is the symmetric decoder; this decoder 
associates the measured syndrome with the error that acts on the fewest number of 
qubits and is consistent with the syndrome. If multiple errors acting on equal 
numbers of qubits are consistent with a syndrome, one is chosen arbitrarily and
used each time that syndrome occurs. 
However the particular choice could affect the threshold value.

The symmetric decoder for the \codepar{5,1,3} code, for example, associates 
each 
syndrome to a unique weight-one Pauli operator. Therefore, all weight-one Pauli 
operators are corrected. However, one could choose a different decoder for the 
5-qubit code. If we consider a noise model where only $X$-errors occur, a 
decoder could be chosen which corrects all weight-one and weight-two Pauli $X$ 
errors. However, this decoder would not be able to correct any $Y$ or $Z$ type 
Pauli errors. More details will be provided in 
\cref{sec:AmpPhaseDamp,sec:CoherentErrorThresholdAnalysis}.

\section{Hard decoding algorithm for optimizing error-correcting codes}
\label{sec:OptimizedDecoding}

We now present our optimized hard decoding algorithm that 
determines the 
choice of recovery operators at each level of concatenation. The goal of the
algorithm is to correct the effective noise, that is, to map it to the identity
channel $\mathcal{I}$ as quickly as possible. More formally, let $\epsilon$
be a pre-metric on the space of CPTP maps, that is, a function 
such that $\epsilon(\mathcal{N},\mathcal{M})\geq 0$ with equality
if and only if $\mathcal{N}=\mathcal{M}$ and 
$\epsilon(\mathcal{N},\mathcal{M})=\epsilon(\mathcal{M},\mathcal{N})$
for all CPTP maps $\mathcal{N}$ and $\mathcal{M}$ (this is a pre-metric as
$\epsilon$ does not have to satisfy the triangle inequality). The function 
$\epsilon(\mathcal{N}):=\epsilon(\mathcal{N},\mathcal{I})$ defines an `error 
rate'.

We will set $1-\epsilon(\mathcal{N})$ to be the average gate fidelity to the
identity, defined in \cref{sec:Infidelity}. This choice significantly reduces
the amount of computational resources required to find the optimal recovery 
maps.
The choice of $\epsilon$ may affect the performance of the decoder, however,
we defer an investigation of this to future work.

\begin{figure*}[t!]
	\includegraphics{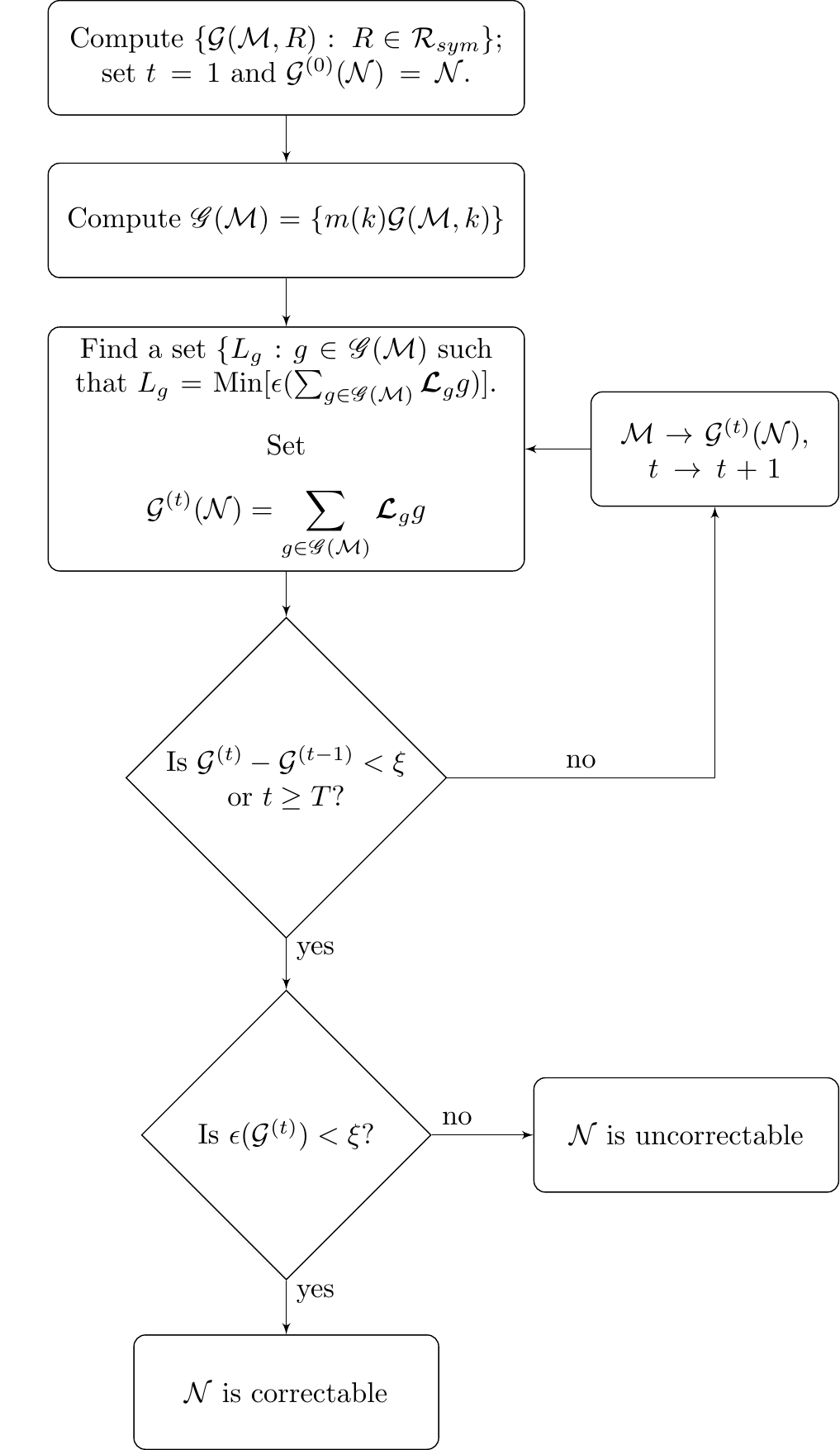}
	\caption{Method for selection of recovery operations for a fixed code $C$ and noise channel $\mathcal{N}$. The iterative step that calculates the process matrix at level $t$, $\mathcal{G}^{(t)}(\mathcal{N})$, is equivalent to setting the recovery maps to $R_{opt} = T(L_g\ad) R$ for all $R\in\mathcal{R}_{sym}$, where $m(k)\boldsymbol{\mathcal{G}}(\mathcal{M},R)=g$ and $T(L_g\ad)$ denotes the transversal implementation of $L_g\ad$. This set may not be unique; see \cref{sec:TieBreaker} for specific examples of when this can occur. We use the $l_{\infty}$ norm, that is, the maximum of the absolute values of the entries of a matrix, to test whether the process matrix has converged. However, any matrix norm can be used instead.}
	\label{fig:RecOpFlow}
\end{figure*}

Our hard decoding optimization algorithm selects recovery operations with the goal of minimizing the logical error rate after the recovery operations have been applied. The flowchart given in \cref{fig:RecOpFlow} applies the hard decoding algorithm to a channel $\mathcal{N}$ and determines whether the effective noise will converge to the identity with concatenation. In \cref{fig:RecOpFlow}, $\mathcal{M}$ is a general CPTP map, $\mathcal{R}_{sym}$ is a (not necessarily unique) set of recovery operators for symmetric decoding (see \cref{sec:SymmetricDecoding}), $\boldsymbol{\mathcal{G}}(\mathcal{M},k)$ are the distinct elements of $\{\boldsymbol{\mathcal{G}}(\mathcal{M},R):R\in\mathcal{R}_{sym}\}$, $m(k)$ is the number of instances of $\boldsymbol{\mathcal{G}}(\mathcal{M},k)$, and $L_g$ is a set of transversal logical operators. The optimized physical recovery maps are
\begin{equation}
	R \to T(L_g\ad)R
	\label{eq:recopsmap}
\end{equation}

or all $R\in\mathcal{R}_{sym}$, where $g=m(k)\mathcal{G}(\mathcal{M},R)$ and $T(L_g\ad)$ denotes the transversal implementation of $L_g\ad$. As we discuss in \cref{sec:NumericalThresholdCal}, the choice for $L_g\ad$ may not be unique. The action of \cref{eq:recopsmap} is equivalent to finding the set $\boldsymbol{\mathcal{L}}_{g} = \{ L_{g} \thinspace : \thinspace g \in \mathscr{G}(\mathcal{M}) \}$ of transversal logical operators that minimize

\begin{align}
\epsilon(\sum_{g \in \mathscr{G}(\mathcal{M})} \boldsymbol{\mathcal{L}}_{g}g)
\label{eq:MinEps}
\end{align}

There are $2^{n-1}$ syndromes, however, step 2 produces only 4, 7, 
12, and 67 distinct process matrices for the \codepar{5,1,3}, Steane, Shor, and 
surface-17 codes respectively, \textit{independently} of the physical noise 
model. Therefore considering only the distinct 
$\boldsymbol{\mathcal{G}}(\mathcal{M},k)$ in step 2 reduces the memory and
computational requirements by a factor between 4 and 20 for the codes considered
in this paper. We could also improve performance by setting the off-diagonal 
terms to zero when they are sufficiently small and recalculating 
$\mathscr{G}(\mathcal{M})$ for Pauli channels $\mathcal{M}$. Removing the 
off-diagonal terms corresponds to performing a Pauli twirl by applying a 
uniformly-random Pauli operator $P$ to each physical qubit before the noise 
acts and then applying a logical $P$ at the $t^{\mathrm{th}}$ concatenation 
level. However, this step complicates the algorithm and was not necessary to 
obtain the results of this paper (it typically sped up computations by a factor 
between 2 and 10).

As we will discuss in \cref{subsec:PauliTwirl}, the use of a code's 
non-Pauli transversal gates (note that for any stabilizer code the logical 
Pauli operators are always transversal) can significantly 
increase performance. This improvement is obtained when a syndrome measurement 
results in a logical 
non-Pauli error with high probability, which can occur even when significantly 
below threshold. This suggests that using highly symmetric stabilizer codes may 
provide better performance even at low error rates (in addition to also making 
non-trivial fault-tolerant computations more viable).   

The resources required to find the set of recovery maps which optimally correct 
a noise model $\mathcal{N}$ are efficient in the number of concatenation levels 
required because our algorithm is independent of the observed syndromes from 
previous concatenation levels. The largest contribution to the complexity of 
our scheme comes from computing the $\beta$ matrix for each syndrome 
measurement. From \cref{eq:BetaIndividualSyn}, there are $3\times2^{n-1}$ 
operations required to compute a $\beta$ matrix for a particular syndrome 
value. The factor of 3 comes from computing the commutation relations (encoded 
by $\eta$) between $R_{l}$ and the code's logical Pauli operators ($R_{l}$ 
always commutes with the identity) and the factor of $2^{n-1}$ comes from 
verifying the commutation relations between $R_{l}$ and all elements in the 
stabilizer group. As there are $2^{n-1}$ possible syndrome values, 
$3\times2^{2(n-1)}$ operations are required to compute all the $\beta$ matrices.

\subsection{Infidelity-optimized decoding}\label{sec:Infidelity}

The average gate infidelity to the identity (hereafter simply the infidelity),
\begin{align}
	r(\mathcal{N}) = 1 - \int \mathrm{d}\psi 
	\bra{\psi}\mathcal{N}(\ket{\psi}\!\bra{\psi})\ket{\psi},
	\label{eq:infidelityConcept}
\end{align}
is a commonly-used error pre-metric on the space of CPTP maps where the 
integral is over all pure states according to the unitarily-invariant 
Fubini-Study metric. The infidelity can be written as
\begin{align}
	r(\mathcal{N}) = \frac{4-\Tr\boldsymbol{\mathcal{N}}}{6}
\end{align}
in the process matrix formalism for a single qubit~\cite{Kimmel2014}. The 
infidelity is particularly convenient for our algorithm because the trace is a 
linear function of the channel. Consequently, to find a set 
$\{L_g:g\in\mathscr{G}(\mathcal{M})\}$ that minimizes 
$\epsilon\Bigl(\sum_{g\in\mathscr{G}(\mathcal{M})} \boldsymbol{\mathcal{L}_{g}}g \Bigl)$ it is sufficient to maximize
\begin{align}\label{eq:MaximizeFidelity}
	\Tr \boldsymbol{\mathcal{L}}g
\end{align}
independently for each $g\in\mathscr{G}(\mathcal{M})$, rather than considering 
all $\lvert \mathscr{G}(\mathcal{M})\rvert^{\lvert \mathscr{L}\rvert}$ 
possibilities.

\subsection{Resolving ties}\label{sec:TieBreaker}

There is one important caveat in the implementation of our hard decoding 
algorithm, namely, there may be multiple sets 
$\{L_g:g\in\mathscr{G}(\mathcal{M})\}$ that minimize the error in 
$\epsilon\Bigl(\sum_{g\in\mathscr{G}(\mathcal{M})} \boldsymbol{\mathcal{L}_{g}}g \Bigl)$. 

For example, consider the Steane code with $\mathcal{U}_\theta(\rho) = U_\theta 
\rho U_\theta\ad$ and $U_\theta = \cos\theta I_2 + i\sin\theta X$. Then the 
only two matrices from step 2 of our algorithm for any value of 
$\theta\in[-\tfrac{\pi}{4},\tfrac{\pi}{4}]$ are
\begin{align}\label{eq:Rot17qubit}
	\boldsymbol{\mathcal{R}}_{z1}(\theta) &= \frac{7 \cos (8 \theta 
		)+25}{32}\boldsymbol{\mathcal{U}_{\phi/2}}\notag\\
	\boldsymbol{\mathcal{R}}_{z2}(\theta) &= \frac{7\sin ^2(4 \theta 
		)}{16}\boldsymbol{\mathcal{U}}_{-3\theta}
\end{align}
for the trivial syndrome and the syndromes that detect $X$ errors 
respectively, where
\begin{align}
	\phi &= \arctan\Bigl(\frac{(3 \cos (4 \theta )+\cos (8 \theta )+10) \tan 
	^3(2 
		\theta )}{-3 \cos (4 \theta )+\cos (8 \theta )+10}\Bigr) \notag\\
	&= 14(\theta^3 + \theta^5) + O(\theta^7).
\end{align}
Similar expressions hold for other values of $\theta$ with different signs. 

For this example, using all transversal gates significantly improves the 
recovery, as, for example, $U_{\pm\pi/4}$ (the phase gate around 
the $X$-axis) is a transversal gate and so 
$\boldsymbol{\mathcal{R}}_{z2}(\pi/12)$ can be perfectly recovered. 
Furthermore, there are two logical gates, namely $U_0$ and $U_{\pi/4}$, that 
maximize \cref{eq:MaximizeFidelity} for $g = 
\boldsymbol{\mathcal{R}}_{z2}(\pi/24)$, and so the choice is ambiguous. When 
confronted with such ambiguities, we choose the first logical operator that 
maximizes \cref{eq:MaximizeFidelity} (in particular, the identity if it is one 
of the options). As we will discuss further in 
\cref{sec:AmpPhaseDamp}, 
this ambiguity due to the ordering of logical operators does arise in practical 
examples without ``fine-tuning'' any parameters and it can impact performance.

\section{Numerically calculating threshold hypersurfaces}
\label{sec:NumericalThresholdCal}

We now describe our numerical method for calculating threshold hypersurfaces 
under symmetric (\cref{sec:SymmetricDecoding}) and infidelity-optimized 
decoders (\cref{sec:OptimizedDecoding}). For convenience, we regard a noise 
channel $\mathcal{N}$ as correctable if there exists some level of 
concatenation $t$ such that $\Tr \boldsymbol{\mathcal{G}}^{(t)}\geq 4-\xi$, or, 
equivalently, the infidelity of $\mathcal{\boldsymbol{G}}^{(t)}$ is at most 
$\xi/6$. The value of $\xi$ was set to 0.01.

\begin{figure*}[t!]
	\includegraphics{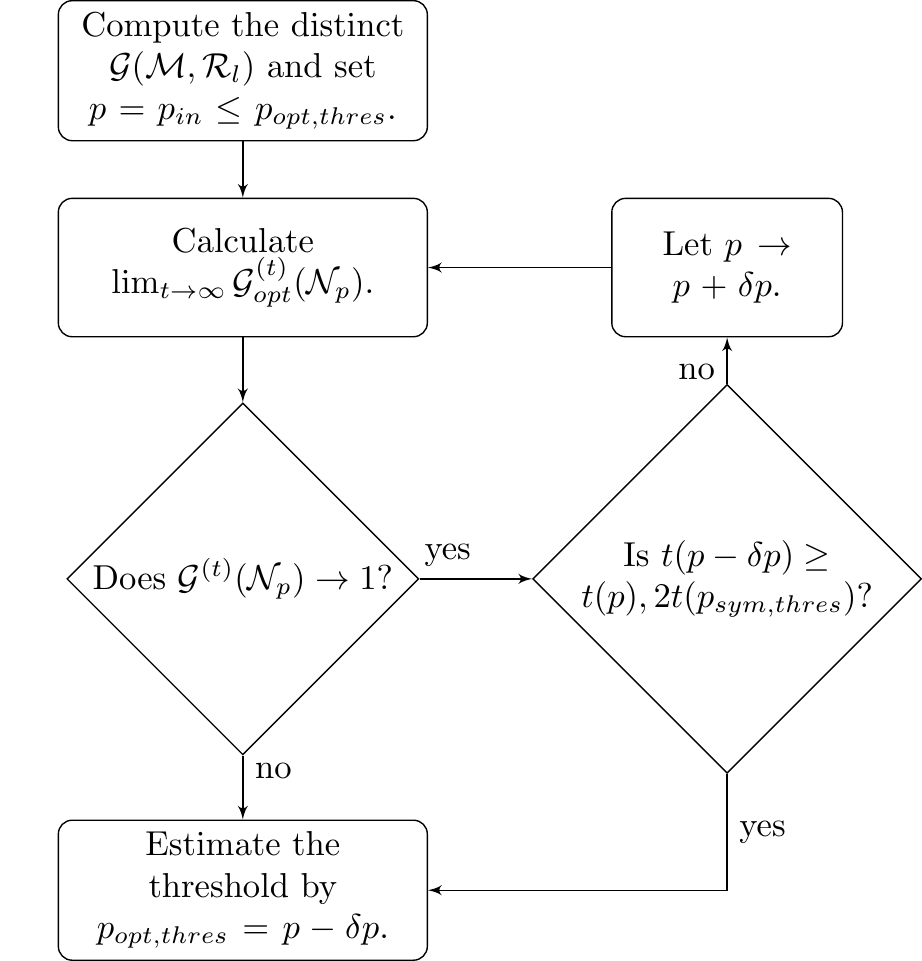}
\caption{Method for lower-bounding the threshold for a one-parameter noise model $\mathcal{N}_p$, where $t(p)$ is the minimum number of concatenation levels required to correct $\mathcal{N}_{p}$. We begin by 
	setting $p_{in} = p_{sym, thres}$ and $\delta p = 0.01$, then 
	repeating with the new lower bound and $\delta p = 0.001$ and finally 
	$\delta p = 0.0001$. To find the threshold hypersurface for an $m$-parameter noise model, repeat this 
	procedure for $\mathcal{N}_{pq}$ while iterating through a mesh of points, $q$.}
\label{fig:chart}
\end{figure*}

\subsection{Symmetric threshold hypersurfaces}

The subset of correctable errors for a given noise model is not generically 
connected. For this reason, a binary search between $p=0$ and $p=1$, where $p$ 
is the noise parameter, is insufficient when calculating a threshold value 
because this method may miss some uncorrectable regimes. To calculate 
$p_{sym,thr}(q)$, a threshold value of $p$ (with $q$ fixed) when a symmetric 
decoder is applied at each level of concatenation, we initialized $p=0$ and 
incremented by 0.05 until the noise with $p=p_u$ was uncorrectable, then 
implemented a binary search between $p=p_u-0.05$ and $p=p_u$. To find a 
threshold hypersurface for a noise model with multiple parameters, we 
iteratively apply this procedure while varying $q$ over a dense mesh.

\subsection{Threshold hypersurfaces for our infidelity-optimized decoder}

To calculate threshold values of a code $C$ afflicted by a general CPTP map using our hard decoding algorithm, we follow the procedure illustrated in \cref{fig:chart}. Here $t(p)$ 
is the minimum number of concatenation levels required to correct 
$\mathcal{N}_{pq}$.

\section{Thresholds and infidelities for amplitude-phase damping}
\label{sec:AmpPhaseDamp}

In the remainder of the paper, we show that our hard decoding algorithm 
leads to significant improvements in threshold values and, in some cases, 
decreases the infidelity by several orders of magnitude relative to the 
symmetric decoder. We will also show that the performance of our decoder is 
robust to perturbations in the noise, so that it can be implemented using 
the necessarily imperfect knowledge of the noise in an experiment.

\begin{figure*}
\begin{center}
\includegraphics[width=.5\textwidth]{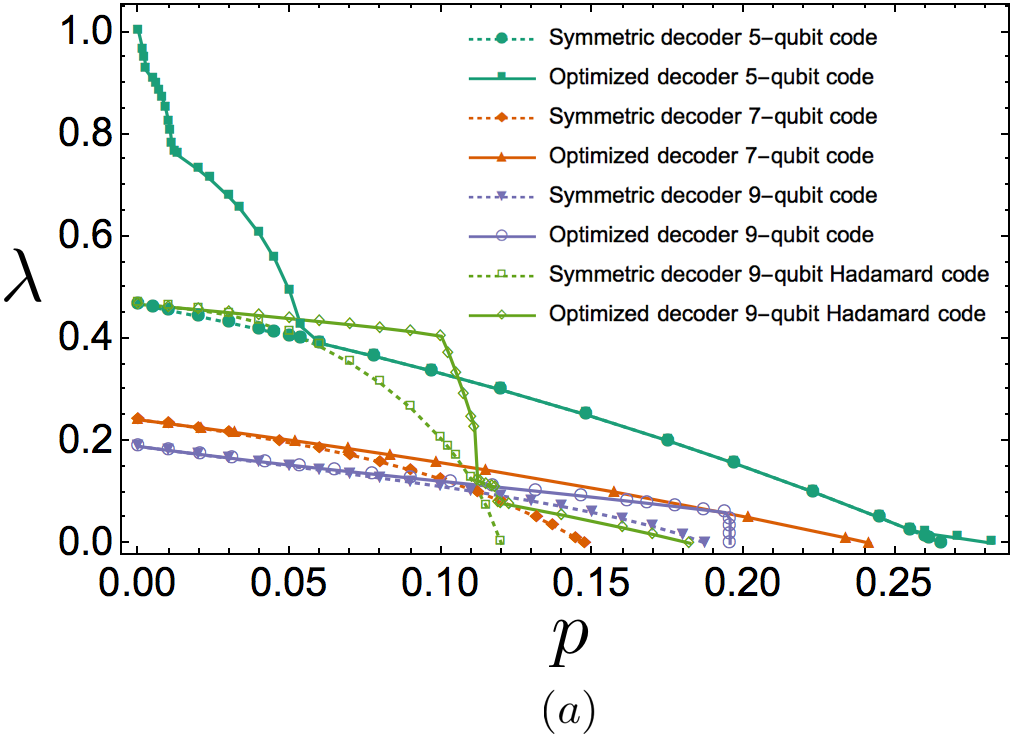}
\end{center}
\begin{minipage}[t]{.5\linewidth}
\begin{center}
\includegraphics[height=55mm]{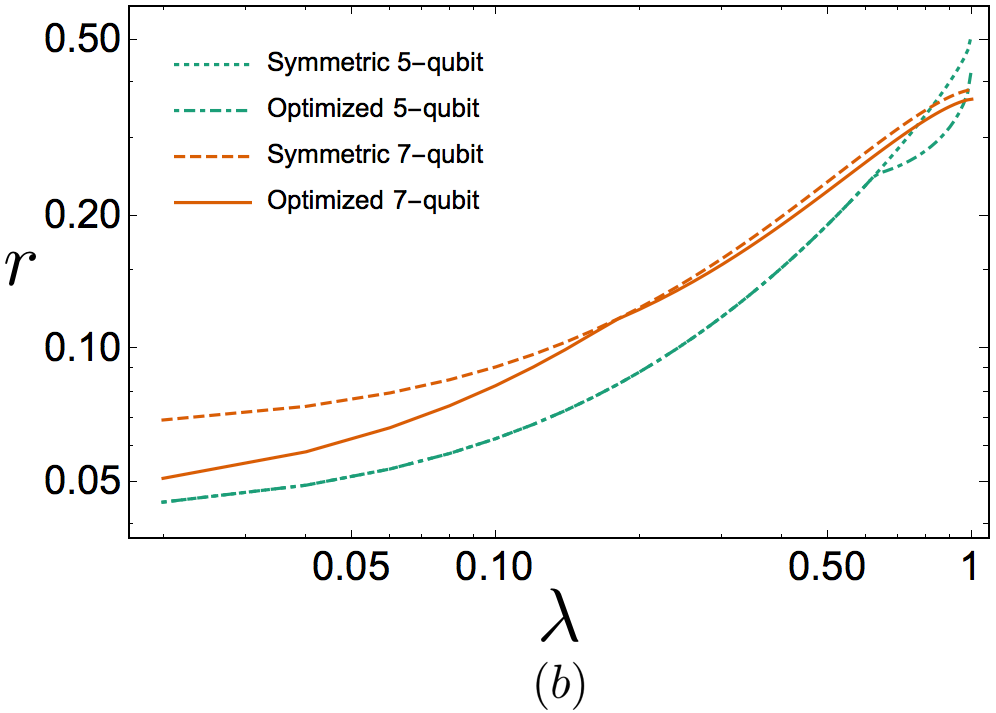}
\includegraphics[height=55mm]{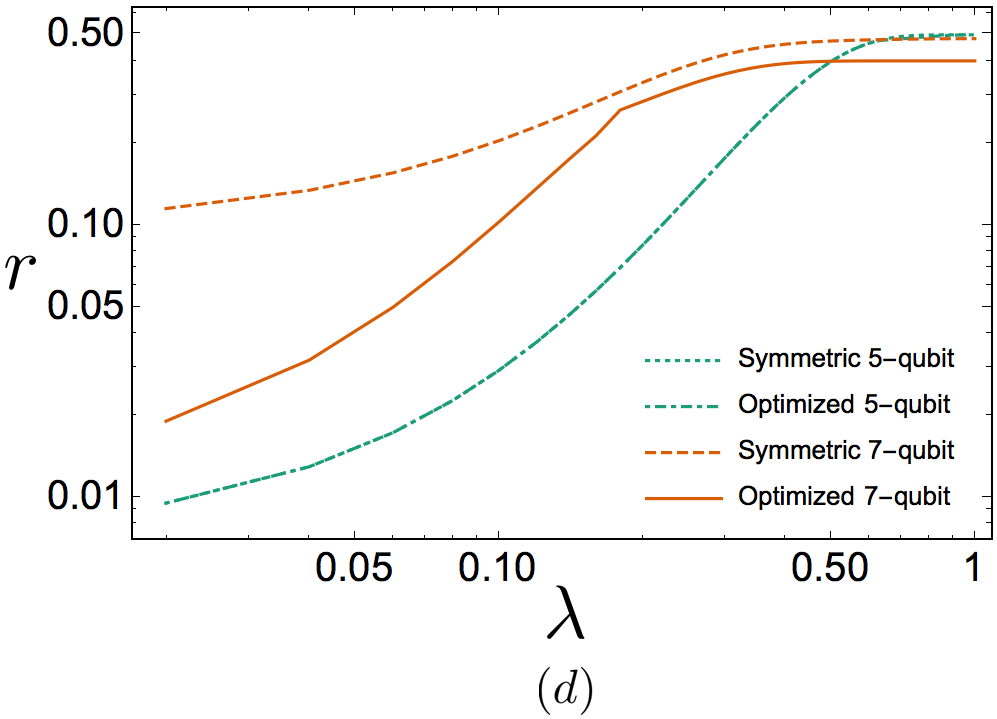}\\
\end{center}
\end{minipage}\hfill
\begin{minipage}[t]{.5\linewidth}
\begin{center}
\includegraphics[height=55mm]{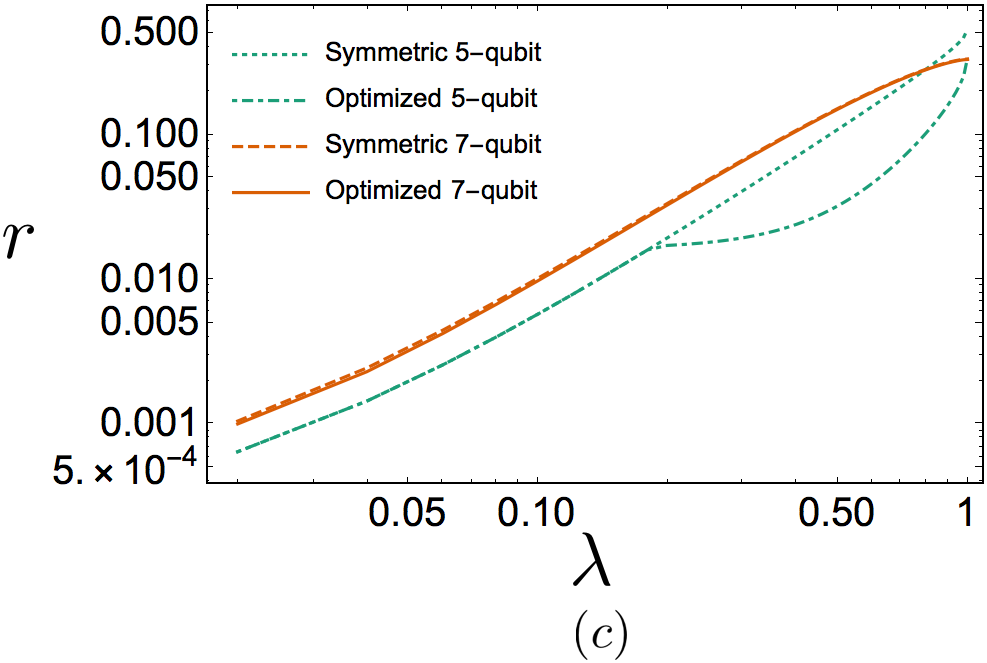}\\
\includegraphics[height=55mm]{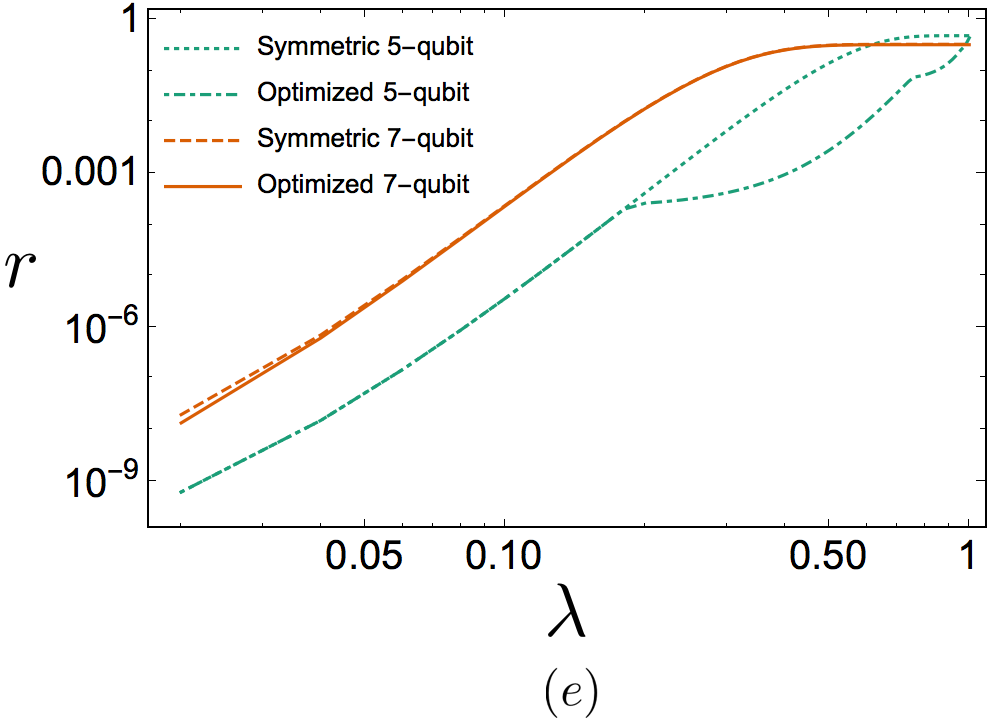}
\end{center}
\end{minipage}
\caption{Threshold curves (a) and infidelities $r$ as functions of the 
dephasing parameter $\lambda$ at the first (b), (c) and third (d), (e) concatenation level for the 
amplitude-phase damping channel under the \codepar{5,1,3} and Steane codes. The Shor and surface-17 codes behave similarly to the Steane code and so their infidelity curves are not displayed. Our numerics also show that for small $p$, the 5 and $X$-Shor codes have the highest thresholds due to lowest number of qubits and asymmetry of stabilizers matching the asymmetry of the noise. The amplitude damping rate is fixed at $p = 0.17$ (b), (d) and $p=0.01$ (c), (e). The symmetrized and optimized curves overlap when the optimized decoder is the symmetric decoder, however, there are many regimes where the optimized decoder improves the threshold and infidelity. For the \codepar{5,1,3} code, thresholds 
using the optimized decoder increase by as much as a factor of 2.14 relative to 
the symmetric decoder. Infidelities are lowered by as much as 2 orders of 
magnitude. The curves for the fixed symmetric 
decoders are all smooth, whereas the curves for the optimized decoders have 
kinks corresponding to points where the decoder changes to exploit asymmetries 
in the noise. When $\lambda$ is large compared to $p$, the noise is primarily dephasing. 
The \codepar{5,1,3} code can exploit this by correcting all single- and two-qubit $Z$ 
errors for the first $t$ concatenation levels, until the noise becomes unbiased 
(the kink in (b) corresponds to $t=1$, and the two in (c) correspond to 
$t=2,3$). The $X$-Shor code can also be biased to correct Pauli-$Z$ errors as it 
has more $X$-type stabilizers. All codes also exhibit improved performance for 
large values of $p$ relative to $\lambda$, that is, when the noise is primarily 
amplitude damping.}
\label{fig:AmpDampPlot}
\end{figure*}

In this section we consider a physical noise model consisting of both amplitude 
and phase damping processes. The amplitude damping channel acts on a two-level 
system at zero temperature. If the system is in the excited state, then a 
transition to the ground state occurs with probability $p$. If the system starts 
in the ground state, it will remain in the ground state indefinitely. A physical
example of this scenario would be the spontaneous emission of a photon in a 
two-level atom. The Kraus operators for the amplitude damping channel are 
\cite{Nielsen_Chuang}
\begin{align}
A^{(0)}_{AD}=\left( \begin{array}{cc}
  1 & 0\\
  0 & \sqrt{1-p}\\                             
  \end{array} \right), \hspace{0.1cm} A^{(1)}_{AD}= \left( 
  \begin{array}{cc}                                                             
    0 & \sqrt{p}\\
	0 & 0\\ \end{array} \right).
\label{eq:KrausOpsAmpDamp}
\end{align}

We point out that \cref{eq:KrausOpsAmpDamp} can be generalized to take into account non-zero temperature effects. In this case, when the system is in the ground state, there is a non-zero probability of making a transition to the excited state. In \cite{CP14}, the performance of the 5-qubit code, Steane code and non-additive quantum codes was estimated for the generalized amplitude damping channel. However, the methods used did not allow for an exact analysis. In the remainder of this manuscript we will only consider the amplitude-damping channel at zero temperature.

Phase damping arises when a phase kick $\exp(i\theta Z)$ is applied to a qubit 
with a random angle $\theta$. When $\theta$ is sampled from a Gaussian 
distribution, then the Kraus operators are
\begin{align}
A^{(0)}_{PD}=\left( \begin{array}{cc}
	1 & 0\\
	0 & \sqrt{1-\lambda}\\                             
\end{array} \right), \hspace{0.1cm} A^{(1)}_{PD}= \left( \begin{array}{cc}
	0 &0\\
	0 &  \sqrt{\lambda}\\                             
\end{array} \right),
\label{eq:KrausOpsPhaseDamp}
\end{align}
where $\lambda$ characterizes the width of the distribution of $\theta$. The 
phase damping channel is also equivalent to the phase-flip channel, that is, 
applying a $Z$ with probability $\alpha = (1 + \sqrt{1 - \lambda})/2$.

Combining the amplitude and phase damping channel, we consider the 
amplitude-phase damping channel given by
\begin{align}
\mathcal{N}_{APD}(\rho) = \mathcal{N}_{PD}(\mathcal{N}_{AD}(\rho))=\mathcal{N}_{AD}(\mathcal{N}_{PD}(\rho)).
\label{eq:AmpPhaseDampChan}
\end{align}
As the amplitude-phase damping channel contains two parameters ($p$ and 
$\lambda$), the threshold hypersurface will be a curve below which the process 
matrix is correctable.  

The threshold curves and infidelity at the first and third concatenation 
levels for the \codepar{5,1,3}, Steane, and Shor codes are illustrated in 
\cref{fig:AmpDampPlot}. The infidelities for the Shor and surface-17 codes are not shown since they behave similarly to the Steane code. The \codepar{5,1,3} code generally outperforms all other codes in terms of
logical infidelity and thresholds under both optimized and symmetric 
decoders, except in an intermediate regime where the optimized decoder exploits
the asymmetry in the stabilizers of the $X$-Shor code. 

The optimized decoder coincides with the symmetric decoder for each code in some
parameter regimes, although only when $p=0$ for the Steane and Shor codes. 
However, the optimized decoder often differs significantly from the symmetric
decoder, resulting in substantially improved logical infidelities and 
thresholds. The optimized decoder changes in different parameter regimes to
exploit asymmetries in the noise, producing the kinks in the curves in 
\cref{fig:AmpDampPlot}. The amplitude-phase damping channel is highly biased
towards $Z$ errors for small values of $p$ relative to $\lambda$. The optimized
decoder exploits this for the \codepar{5,1,3} code by only correcting $Z$ errors
for $t$ levels of concatenation until the noise is approximately symmetric, and
then switching to the symmetric decoder, with $t$ increasing as $p$ approaches
zero. The $X$-Shor code also performs better in this regime as it has more 
$X$-type stabilizers that detect $Z$-type errors.

The optimized decoder also results in improved thresholds and logical 
infidelities for high amplitude damping rates for all codes. The noise is 
significantly different from Pauli noise in this regime and so decoders 
constructed under the assumption of Pauli noise will be less likely to identify 
the correct error compared to decoders optimized for amplitude-phase damping.

As discussed in \cref{sec:TieBreaker}, multiple sets 
$\{L_g:g\in\mathscr{G}(\mathcal{M})\}$ maximize \cref{eq:MaximizeFidelity} for
the Steane code with amplitude-phase damping and large values of $\lambda$. 
For example, setting $\lambda=0.1431$ and choosing the first recovery operator 
that maximizes \cref{eq:MaximizeFidelity} gives a threshold of $p_{th}= 0.1032$. 
However, searching all tuples $\{L_g:g\in\mathscr{G}(\mathcal{M})\}$ that 
maximize \cref{eq:MaximizeFidelity} (where the degeneracy only occurs at the 
first level) for the same value of $\lambda$ gives a higher threshold of 
$p_{th} = 0.1150$.

\section{Thresholds for coherent errors}
\label{sec:CoherentErrorThresholdAnalysis}

\begin{figure*}[t!]
\begin{subfigure}{0.33\textwidth}
\begin{center}
\includegraphics[width=\textwidth]{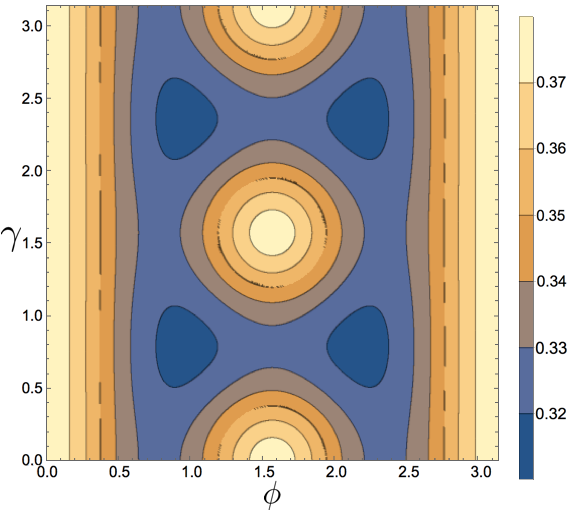}
\caption{}
\includegraphics[width=\textwidth]{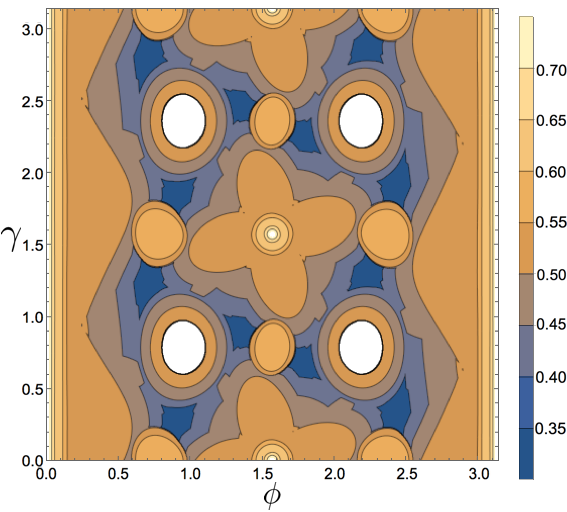}
\caption{}
\end{center}
\end{subfigure}\hfill
\begin{subfigure}{0.33\textwidth}
\includegraphics[width=\textwidth]{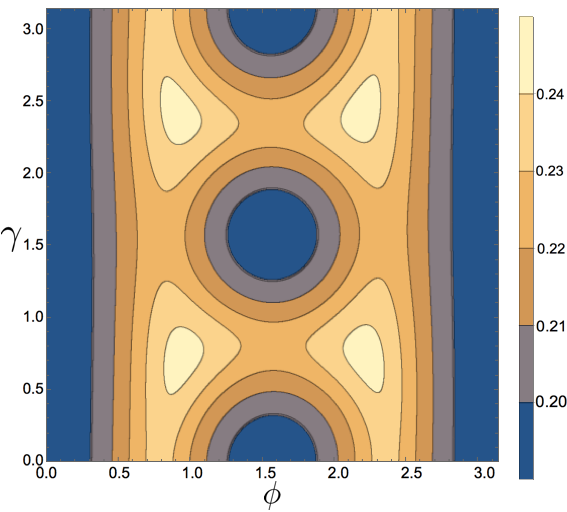}
\caption{}
\includegraphics[width=\textwidth]{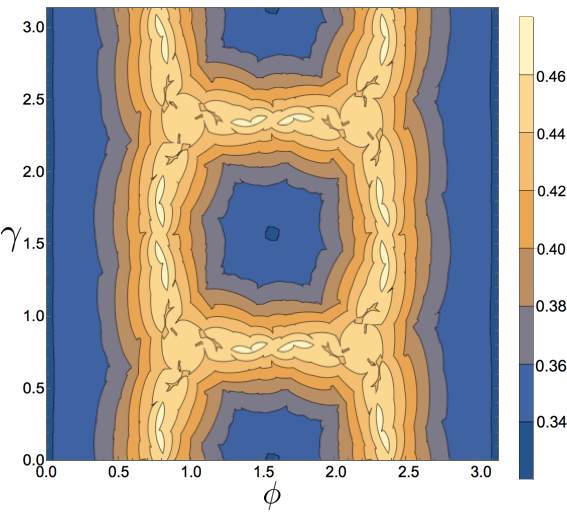}
\caption{}
\end{subfigure}\hfill
\begin{subfigure}{0.33\textwidth}
\includegraphics[width=\textwidth]{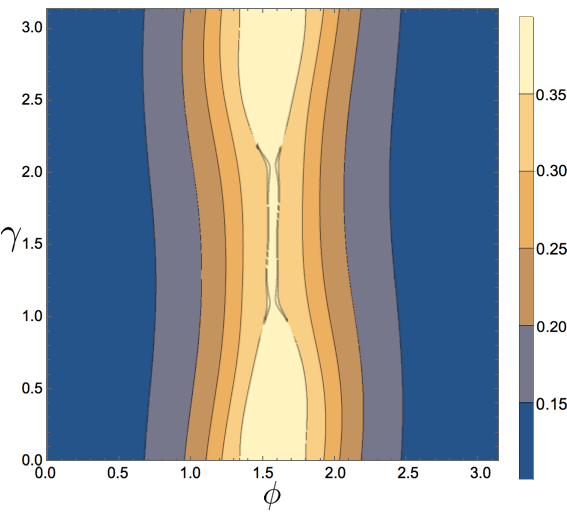}
\caption{}
\includegraphics[width=\textwidth]{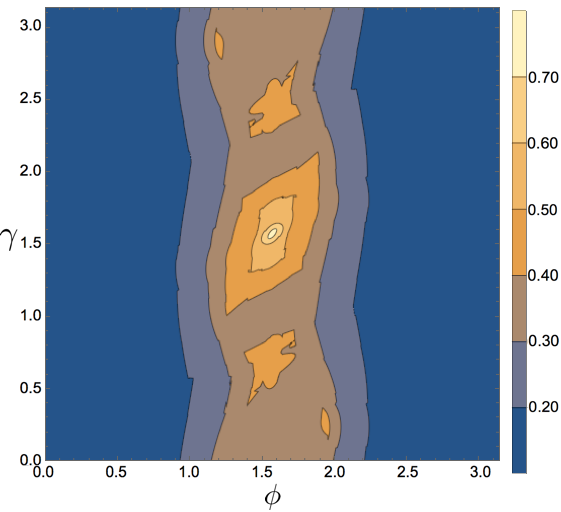}
\caption{}
\end{subfigure}\hfill
\caption{Contour plots representing hypersurfaces of the threshold value of $\theta$ for rotations around
the axis $\hat{n} = (\sin\phi\cos\gamma,\sin\phi\sin\gamma,\cos\phi)$ for (a) the
\codepar{5,1,3} code, (c) the Steane code and (e) the Shor code using the symmetric decoder and (b) the \codepar{5,1,3} code, (d) the Steane code and (f) the Shor code using optimized decoding.
The optimized decoder uses transversal gates to improve the
threshold, particularly when the rotation is around an eigenbasis of a 
transversal Clifford gate (see \cref{tab:StabilizerGeneratorsLists}). In 
particular, the \codepar{5,1,3} code has a transversal $\pi/3$ rotation around
$\hat{n}_C = (1,1,1)/\sqrt{3}$ (i.e., $\gamma = \pi/4$, $\phi = \pi/3$), which
enables the optimized decoder to correct \textit{arbitrary} rotations around axes
close to $\hat{n}_C$, illustrated by the white circular regions in (b). For Steane's 
code, the lightest colored regions in (d) corresponds to threshold angles 
$\theta_{th} \approx 0.46$ compared to $\theta_{th} \approx 0.24$ in (c), an 
improvement by almost a factor of 2. The Shor code has no 
transversal non-Pauli gates and so the improvements from the optimized decoder 
are not as substantial. However, for rotations near the $y$-axis, the Shor code outperforms the Steane code by a factor of at most 2.3.}
\label{fig:CoherentPlot}
\end{figure*}

\begin{figure*}[t!]
\begin{subfigure}{0.50\textwidth}
\begin{center}
\includegraphics[width=\textwidth]{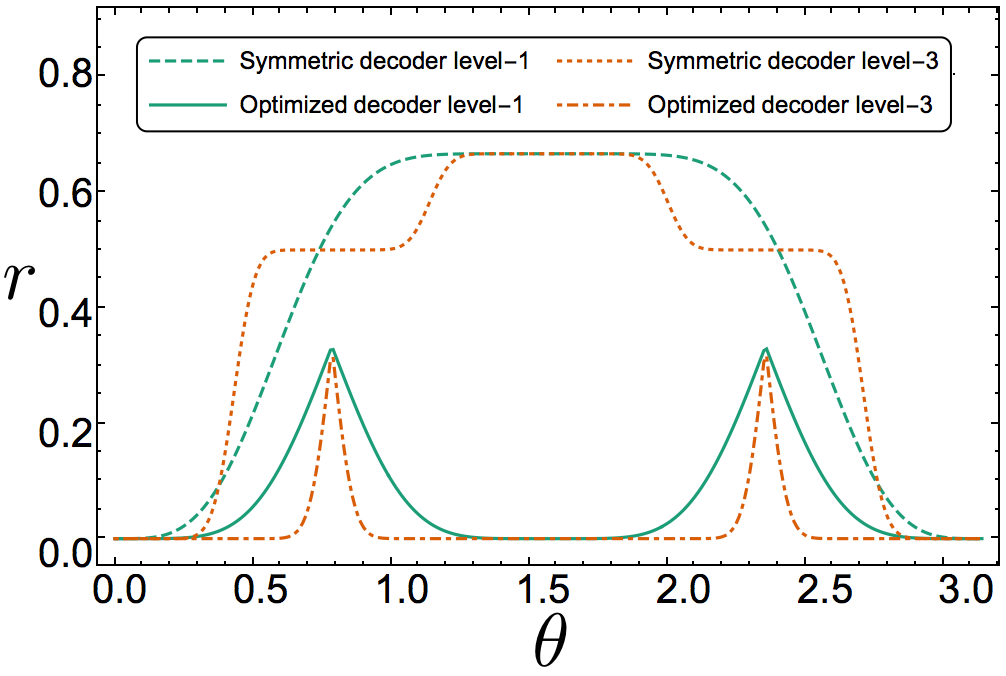}
\caption{}
\includegraphics[width=\textwidth]{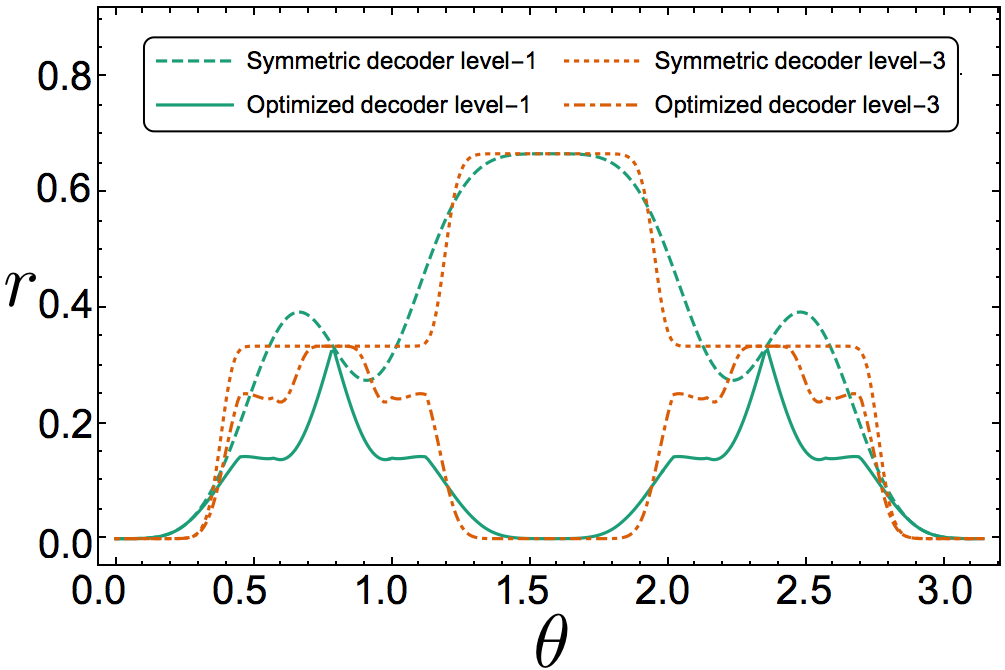}
\caption{}
\end{center}
\end{subfigure}\hfill
\begin{subfigure}{0.50\textwidth}
\begin{center}
\includegraphics[width=\textwidth]{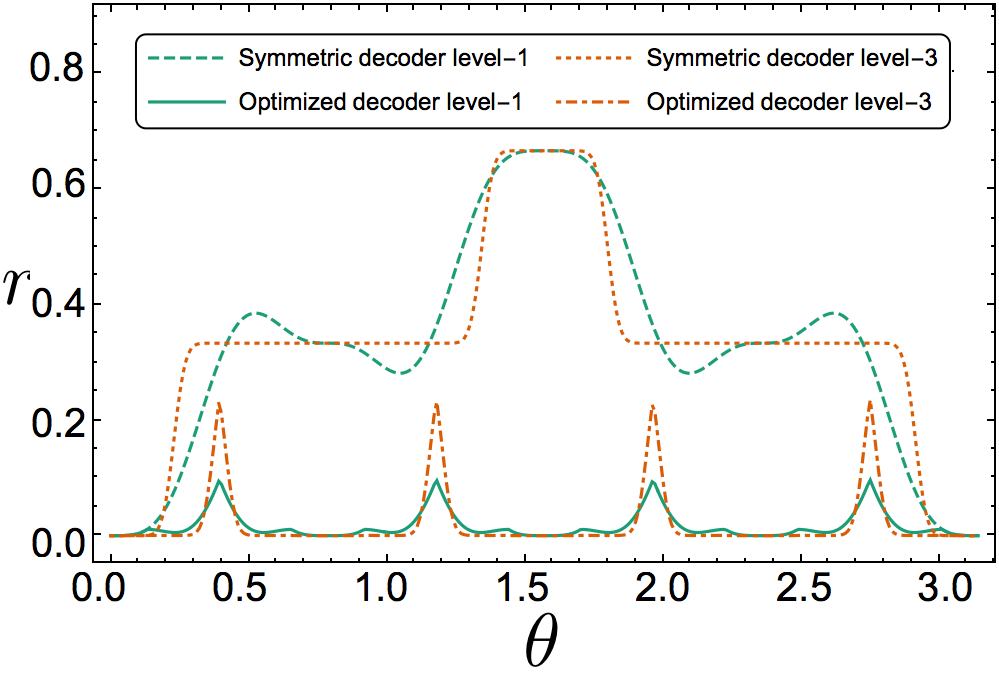}
\caption{}
\includegraphics[width=\textwidth]{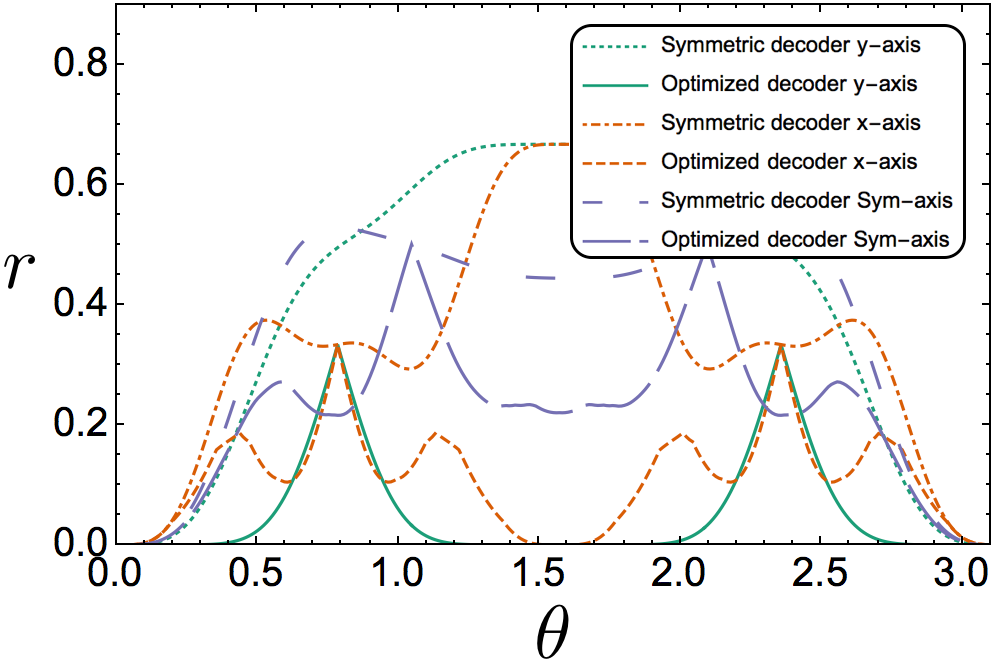}
\caption{}
\end{center}
\end{subfigure}\hfill
\caption{In (a), (b) and (c) infidelities $r$ of the \codepar{5,1,3} code, 
Shor code and Steane code are plotted at the first and third levels for a 
rotation about the $x$-axis. In (a), the optimized infidelity curves are peaked 
at the code's threshold value $\theta_{th} = \pi/4$. In (b), the peaks of the 
optimized infidelity curves are centred slightly above the code's threshold 
value $\theta_{th} =0.3396$. However, the optimized level-3 infidelity curve 
intersects the optimized level-1 curve at the threshold value as expected. In 
(c), the optimized level-1 and level-3 infidelity curves intersect at the threshold value $\theta_{th} = 0.3692$. The peaks of the infidelity curves occur at $\theta = \pi/4$ due to the codes symmetry. In 
(d), infidelity plots of the surface-17 at the first concatenation level are 
shown for a rotation about the $y$-axis, $x$-axis and the $(1,1,1)/\sqrt{3}$ 
axis. It can be seen that the infidelity is lowest for rotations about the 
$y$-axis. In all 4 plots, it can be seen that applying our hard decoding 
algorithm reduces the infidelities by, in some cases, orders of magnitude compared to the symmetric decoder.}
\label{fig:CoherentInfidelities}
\end{figure*}

In this section we illustrate the behavior of coherent errors under error 
correction. We consider a coherent error noise model where every qubit 
undergoes a rotation by an unknown angle $\theta$ about an axis of rotation 
$\hat{n}$. The coherent noise channel can thus be written as
\begin{align}
\mathcal{N}_{\theta,\phi,\gamma}(\rho) =e^{i \theta \hat{n} \cdot \vec{\sigma}} \rho 
e^{-i \theta \hat{n} \cdot \vec{\sigma}},
\label{eq:CoherentErrorChan}
\end{align}
where $\hat{n}=(\sin\phi\cos\gamma,\sin\phi\sin\gamma,\cos\phi)$. We obtain the 
threshold
hypersurface for the noise model 
\begin{align}
\mathscr{N} = \{\mathcal{N}_{\theta,\phi,\gamma}:\theta\in[0,2\pi],\gamma,\phi\in[0,\pi]\},
\end{align}
by fixing $\gamma$ and $\phi$ and obtaining the threshold for $\theta$.
 
The threshold hypersurfaces for the \codepar{5,1,3}, Steane, and Shor codes are 
illustrated as contour plots in \cref{fig:CoherentPlot}. The infidelities at
the first and third concatenation levels for \codepar{5,1,3}, Steane, Shor, and
surface-17 (1st level only) codes are plotted in 
\cref{fig:CoherentInfidelities}.

Unlike with amplitude-phase damping noise, the optimized decoder strictly 
outperforms the symmetric decoder for all rotation axes. With the exception of
the Shor code, the threshold hypersurfaces are relatively flat under symmetric
decoding, that is, the threshold rotation angle is relatively independent of the
rotation axis. The optimized decoder breaks this, giving larger threshold 
angles for different axes, particularly for rotations about an eigenbasis of 
the transversal Clifford gates listed in \cref{tab:StabilizerGeneratorsLists}. 
In particular, the \codepar{5,1,3} code can correct any rotation about axes 
$\hat{n}$ that are close to $(\pm,\pm,\pm)/\sqrt{3}$.
The Steane code can correct any rotation about 
the Pauli axes except for angles close to odd integer multiples of $\pi/4$. 
The performance of the Shor code is generally only modestly
improved by the optimized decoder, largely because the Shor code has
no transversal non-Pauli gates. However, the optimized decoder is able to 
exploit the asymmetries in the stabilizer generators to increase
the threshold for rotations near the $y$ axis by more than a factor of 3.

The improved threshold angles are reflected in the 
orders-of-magnitude reduction in the logical infidelities in 
\cref{fig:CoherentInfidelities}. The infidelities are periodic because 
transversal gates can be used to counteract the unitary noise. However, as
discussed in \cref{subsec:PauliTwirl}, the transversal gates are useful even 
when the action of the noise on the codespace is far from a transversal gate. For 
the \codepar{5,1,3}, Steane and $Z$-Shor codes, the infidelities are the 
infidelity 
at the first and third concatenation levels for rotations about the $x$ axis, 
while for the surface-17 code the infidelities are at the first level for 
rotations about the $x$, $y$ and $\frac{1}{\sqrt{3}}(1,1,1)$ axes. For the \codepar{5,1,3}, Steane and Shor codes, the threshold values of $\theta$ correspond exactly to the cross-over points 
between the level-1 and level-3 curves. At the third level, the infidelity is greatly suppressed below threshold 
and increased above threshold. The optimized infidelity curves are lower (in 
some cases by several 
orders of magnitude) than the infidelities arising by applying the symmetric 
decoder at all levels. 

For the \codepar{5,1,3} code, the only uncorrectable values of 
$\theta$ are odd integer multiples of $\pi/4$. The surface-17 code has a higher threshold against $Y$ 
errors than against $X$ (or $Z$) errors. The surface-17 code treats $X$ and $Z$ 
errors symmetrically. However, since the $X$ and $Z$ stabilizer generators have 
support on different qubits, error rates resulting from $Y$ errors will differ 
from error rates resulting from $X$ and $Z$ errors. 

The optimized decoding algorithm gives the greatest improvements for codes with 
transversal non-Pauli gates, namely, the \codepar{5,1,3} and Steane codes.

\section{Correlated noise channel}
\label{sec:CorrelatedNoiseThresholdAnalysis}

\begin{figure*}[t!]%[htbp]
\begin{subfigure}{0.48\textwidth}
\begin{center}
\includegraphics[height=55mm]{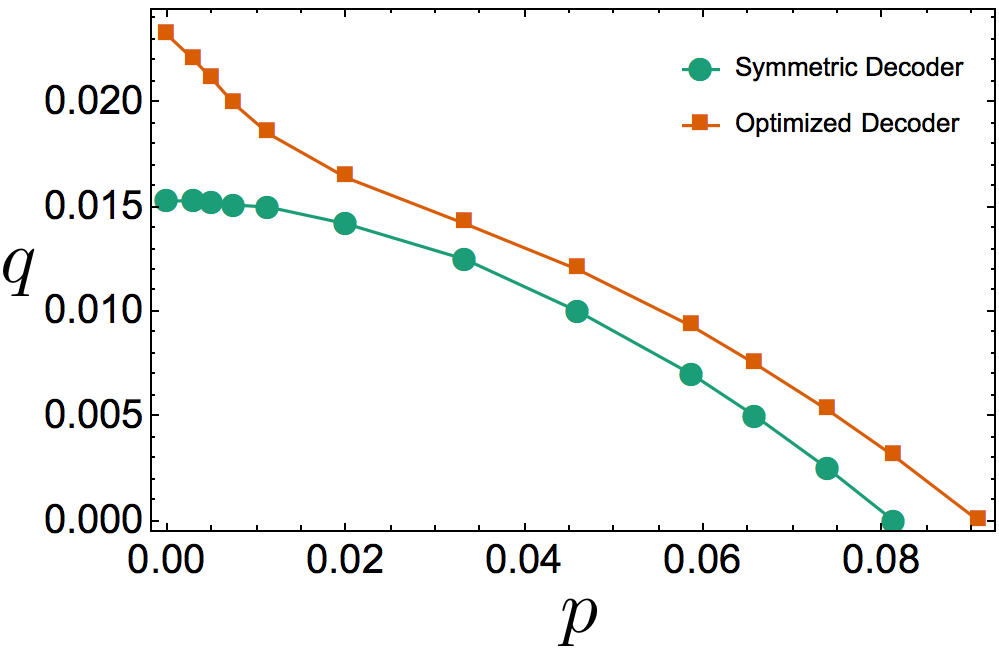}
\caption{}
\end{center}
\end{subfigure}\hfill
\begin{subfigure}{0.48\textwidth}
\begin{center}
\includegraphics[height=55mm]{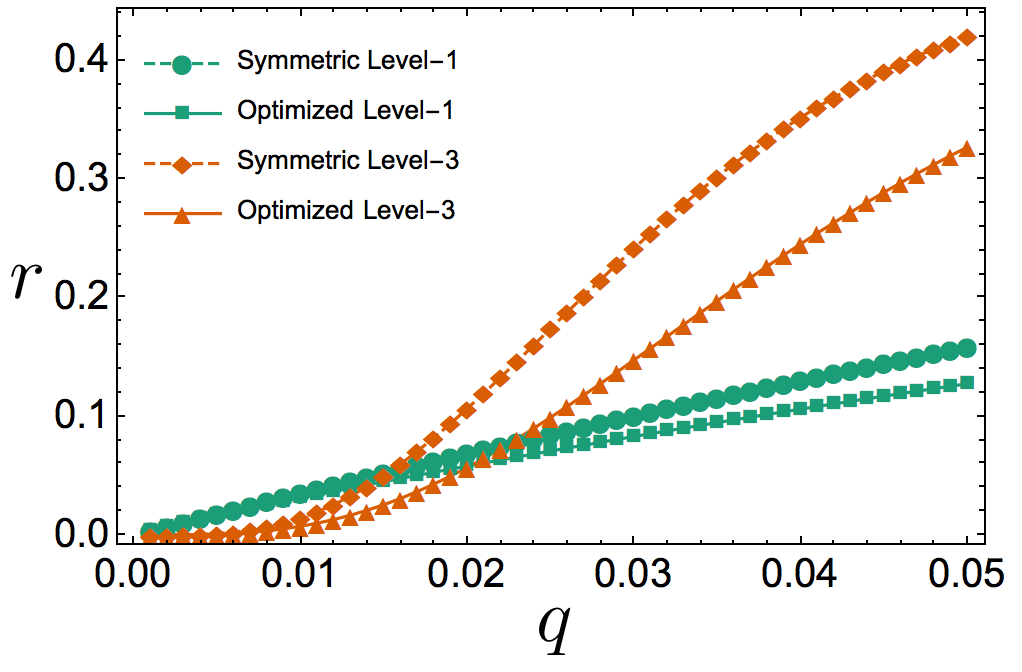}
\caption{}
\end{center}
\end{subfigure}\hfill
\caption{ (a) Threshold curves and (b) infidelities $r$ at the first and third 
concatenation level (with fixed $p=0.003$) for the \codepar{7,1,3} code. 
Two-qubit correlated dephasing occurs with probability $q$ and depolarizing 
noise occurs with probability $1-q$, with a depolarizing noise parameter $p$ 
(see \cref{eq:CorrelatedNoise2,eq:DepolarzingContribCorrelated}). For small 
values of $p$, the $Z$ errors arising from the two-qubit correlations dominate 
the noise. Applying our optimized hard decoding algorithm in this regime yields 
a threshold of $q_{th} = 0.0232$. The contribution from depolarizing noise
increases with $p$ until the noise is predominantly depolarizing. In this 
regime, the optimized decoder implements the standard CSS decoder at all levels.
When $q=0$, the noise is purely depolarizing and $p_{th}=0.0908$. For all 
values of $p$, the threshold $q_{th}$ obtained by implementing our optimized 
decoder is larger than the threshold obtained by implementing the symmetric 
decoder. The level-1 
and level-3 infidelity curves intersect near the respective thresholds for 
$p=0.003$, namely, $q_{th}=0.0153$ and $q_{th} = 0.0220$ for the symmetric and
optimized decoders respectively.}
\label{fig:CorrelatedNoisePlots}
\end{figure*}

In this section, we study the effect of correlated noise on the logical noise
in Steane's code. The correlated noise we consider consists of local 
depolarizing noise and two-qubit correlated dephasing errors to all adjacent 
pairs. The composite noise channel maps an $n$-qubit state $\rho$
\begin{align}\label{eq:CorrelatedNoise2}
\mathcal{N}(\rho) = (1-q)\mathcal{D}_p^{\otimes n}(\rho)+ 
\frac{q}{n}\sum_{j=1}^n\mathcal{Z}_{j,j+_n 1}^{(2)}(\rho),
\end{align} 
where $j+_n 1 = j+1$ if $j<n$ and $1$ otherwise (that is, we consider the qubits
to be in a ring),
\begin{align}
\mathcal{D}_p(\tau) = (1-p)\tau + \frac{p}{3}(X \tau X + Y \tau Y + Z \tau Z)
\label{eq:DepolarzingContribCorrelated}
\end{align}
is the depolarizing channel acting on a single-qubit state $\tau$, and 
$\mathcal{Z}_{j,j+_n 1}(\rho)= Z_{j}Z_{j+1}\rho 
Z_{j+1}Z_{j}$ applies phase-flip operators to qubits $j$ and $j+1$. The logical
process matrix can be computed for the noise model in \cref{eq:CorrelatedNoise2}
by \cref{eq:CorrelatedNoiseProcessMat}.

The threshold and infidelity at the first and third concatenation levels of 
Steane's code are illustrated in \cref{fig:CorrelatedNoisePlots}. In the small $p$ regime, the noise is dominated by the two-qubit correlated 
dephasing contribution. The optimized decoder corrects a larger amount of $Z$ errors at the first few levels by breaking the symmetry in the syndrome measurements. At higher levels, the decoder corrects in a more symmetric fashion in order to remove the remaining Pauli errors. 
This improved performance is also 
illustrated in the reduced optimized logical infidelities shown in 
\cref{fig:CorrelatedNoisePlots}~(b) as a function of $q$ with
 $p=0.003$.

There is an intermediate regime where the local depolarizing noise contribution 
becomes more relevant, leading to a decrease in the threshold value for
$q$. However, the optimized threshold is still noticeably larger than the 
symmetric decoder threshold. Finally, when the local depolarizing noise is
the dominant source of noise, our optimization algorithm chooses recovery maps 
consistent with the standard CSS decoder. The standard CSS decoder yields a 
slightly larger $p$ threshold value compared to the symmetric decoder when 
$q=0$.

\section{The effect of Pauli twirling on thresholds and the benefits of using 
transversal operations}
\label{subsec:PauliTwirl}

In 
\cref{sec:AmpPhaseDamp,sec:CorrelatedNoiseThresholdAnalysis,sec:CoherentErrorThresholdAnalysis},
 we showed that our hard decoding optimization algorithm could  
improve threshold values by more than a factor of 2 for amplitude-phase damping 
noise. For coherent noise there where certain rotation axes where the noise was 
correctable for arbitrary rotation angles. Infidelities were reduced by orders 
of magnitudes in certain regimes. The amplitude-phase damping and coherent 
noise models are both non-Pauli. 
Performing a Pauli twirl on a noise channel $\mathcal{N}$ (that is, conjugating 
it by a uniformly random Pauli channel) maps it to a channel 
$\mathcal{T}(\mathcal{N})$ that is a Pauli channel and so has a diagonal matrix 
representation with respect to the Pauli basis \cite{Emerson2007,Geller2013}. 
In 
\cite{Geller2013,Gutierrez2016}, the effective noise at the first level for the 
amplitude damping 
channel was found to be in good agreement to a Pauli twirled approximation of 
the channel.  

\begin{figure}[h]
\centering
\includegraphics[width=0.45\textwidth]{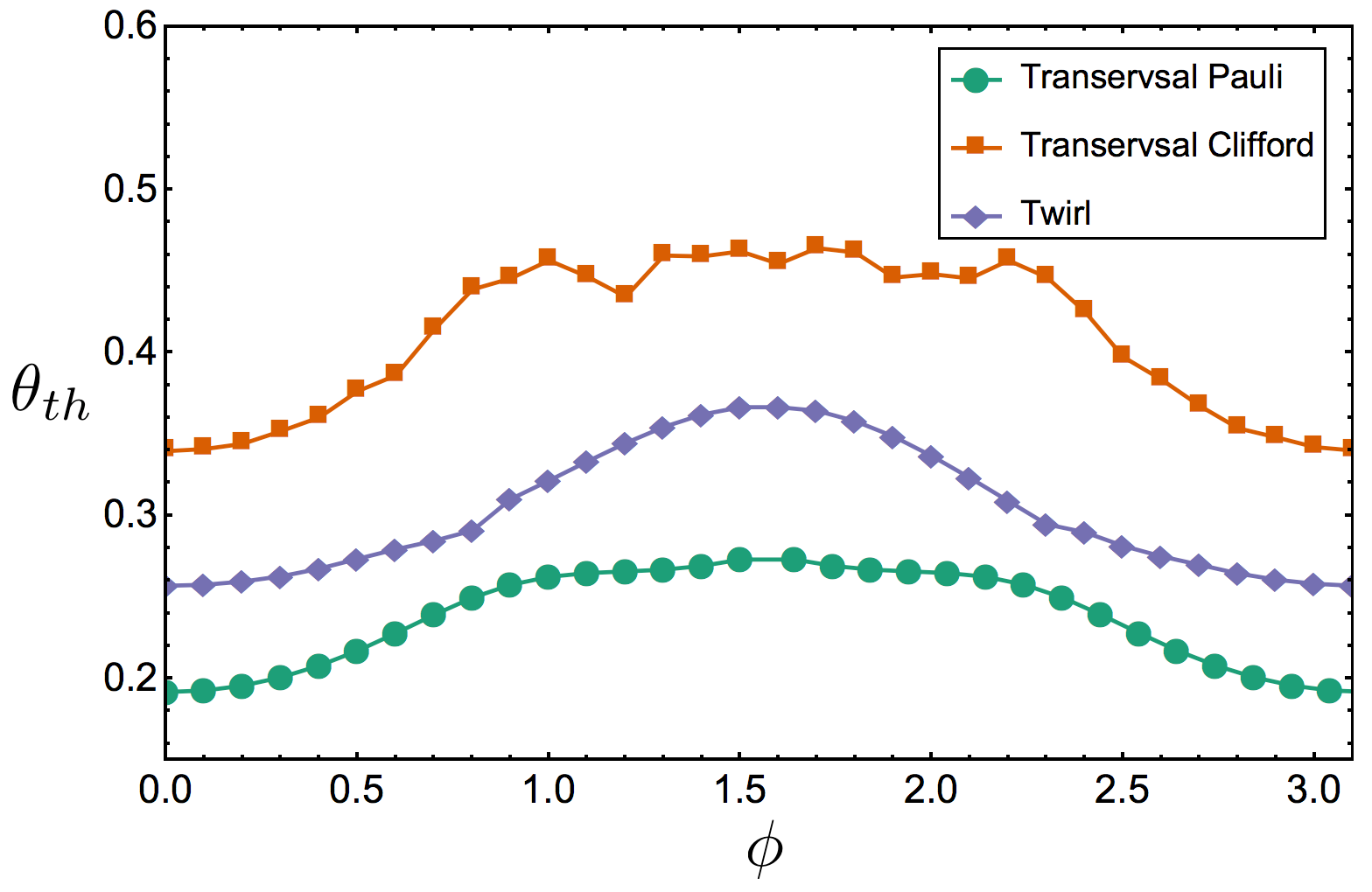}
\caption{Threshold curves for unitary rotations about 
$(\tfrac{1}{\sqrt{2}}\sin\phi,\tfrac{1}{\sqrt{2}}\sin\phi,\cos\phi)$ by an 
angle $\theta$ under three different decoding schemes for the Steane code. For 
all values of $\phi$, an improvement by as much as a factor of 1.7 in the 
threshold $\theta_{th}$ obtained by using our algorithm optimizing over all 
transversal Clifford gates can be observed relative to optimizing 
over all Pauli gates. The Pauli-twirl reduces (increases) the threshold when 
optimizing over all transversal Clifford (Pauli) gates for all values of 
$\phi$.}
\label{fig:7qubitTwirlPlots}
\end{figure}

However, we now show that performing a Pauli twirl on coherent noise and using 
the Steane code can either reduce or increase threshold values, 
depending on the particular recovery protocol. We also illustrate the 
improvements obtained by using all transversal gates in the 
decoding algorithm, instead of just the Pauli gates. Threshold curves for 
rotations about 
$(\tfrac{1}{\sqrt{2}}\sin\phi,\tfrac{1}{\sqrt{2}}\sin\phi,\cos\phi)$ by 
an angle $\theta$ under three different decoding schemes for the Steane code 
are presented in \cref{fig:7qubitTwirlPlots}. The three schemes we consider 
are :
1) our optimized decoding algorithm applied to the twirled noise; 
2)  our optimized decoding algorithm applied to the bare 
noise using all transversal gates; and 3) our optimized decoding algorithm 
applied to the bare 
noise using only transversal Pauli gates.
Using transversal Clifford gates in our recovery protocol gives the largest 
threshold values for all values of $\phi$ and so Pauli twirling reduces the 
threshold. However, if only transversal Pauli operators are used, Pauli 
twirling increases the threshold for all values of $\phi$. 

The curves in \cref{fig:7qubitTwirlPlots} also demonstrate that the threshold 
can increase by at most a factor of 1.7 when optimizing over all transversal 
gates for coherent noise compared to optimizing over all transversal gates for 
the twirled channel. This 
advantage arises for two reasons. First, for a known noise model, a transversal 
gate can be applied to map it to another noise model that may be closer to the 
identity. Second, syndrome measurements may map coherent errors closer to 
a non-Pauli unitary. However, both these benefits are lost when the noise is 
twirled because both the physical noise and the noise for each syndrome is 
Pauli noise, which is generally far from any non-Pauli unitary.

\section{Sensitivity and robustness of our hard decoding optimization algorithm 
to perturbations of the noise model}
\label{subsec:SensitivityDecoder}

In \cref{sec:OptimizedDecoding}, we presented a hard decoding algorithm for 
optimizing threshold values of an error correcting code for arbitrary CPTP maps.
Our algorithm can therefore be applied to non-Pauli channels, including more 
realistic noise models that could be present in current experiments. However, 
the noise afflicting an experimental system is only ever approximately known. Nevertheless, we now 
demonstrate that applying the decoder obtained by our algorithm for a fixed
noise channel $\mathcal{N}$ to a perturbed noise model $\mathcal{N}_p$ retains, 
in some cases, improvements in error suppression relative to the symmetric 
decoder.

To study perturbations about a noise channel, let $\mathscr{N} = 
\{\mathcal{N}_p:p\in[0,1]\}$ be a 1-parameter noise model and
\begin{align}\label{eq:PerturbationOfNoiseChannel}
\mathcal{N}_{U,p}(\rho) = [1-f(p)] \mathcal{N}_p + f(p) U\rho U^\dagger
\end{align}
where $U$ is a random unitary and $f:[0,1]\to[0,1]$ is a function such that 
$f(p)\ll \lVert \mathcal{N}_p-\mathcal{I}\rVert$ for any suitable norm (e.g., 
the diamond norm). (The generalization to multi-parameter noise families is
straightforward.) 

\begin{figure*}[t!]
\begin{subfigure}{0.5\textwidth}
\begin{center}
\includegraphics[width=\textwidth]{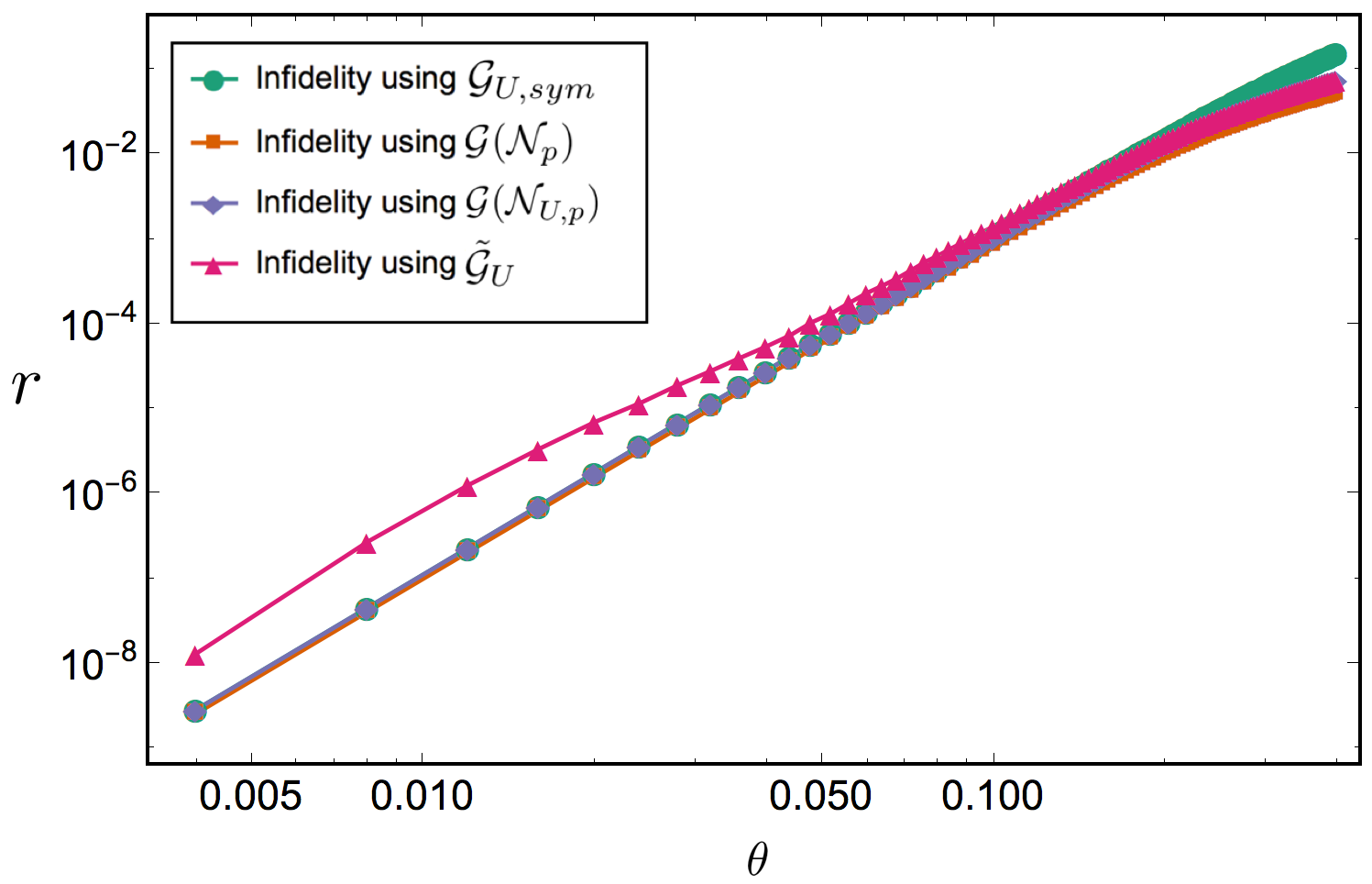}
\caption{}
\end{center}
\end{subfigure}\hfill
\begin{subfigure}{0.5\textwidth}
\begin{center}
\includegraphics[width=\textwidth]{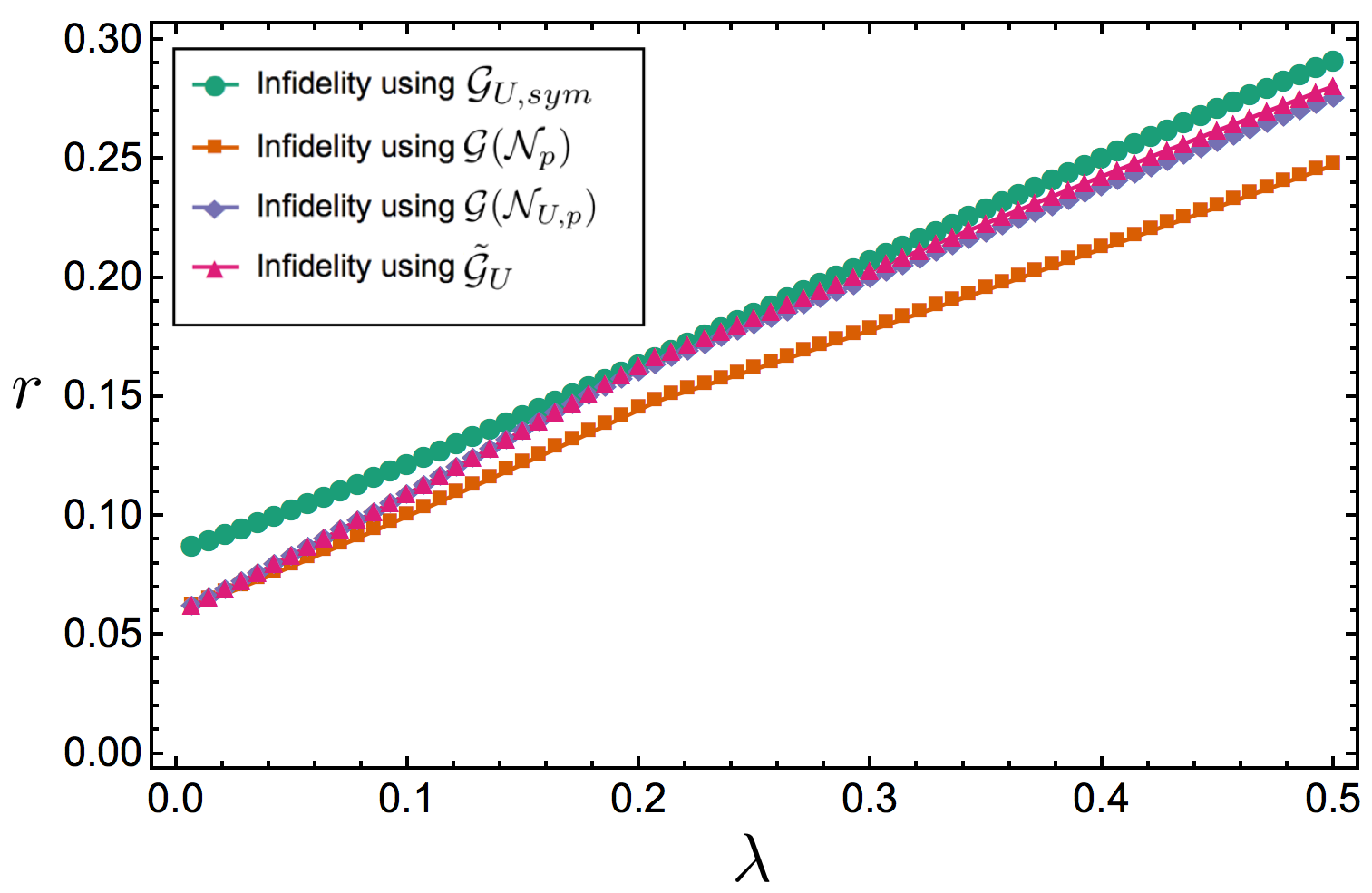}
\caption{}
\end{center}
\end{subfigure}\hfill
\caption{Averaged infidelity plots over 100 random unitary operators $U$ of the effective process matrices $\mathcal{G}(\mathcal{N}_p)$, $\mathcal{G}(\mathcal{N}_{U,p})$,  $\mathcal{G}_{U,sym}$ and $\tilde{\mathcal{G}}_{U}$. The figure in (a) is obtained using the \codepar{5,1,3} code for coherent errors using random rotation axes for each random unitary. The perturbation was chosen to have the form $f(\theta) = \sin^{2}{\theta}/10$. The figure in (b) is obtained using the Steane code for the amplitude-phase damping channel. The perturbation was chosen to have the form $f(\lambda)=\lambda/10$. In (a), the inset plot shows all infidelities on a log-log scale in the regime where $\theta$ is small. As can be seen from the figure, in the regime where $\theta \gtrsim 0.185$, the optimized recovery maps for the unperturbed channel yield a lower infidelity when applied to the perturbed channel than that from applying the symmetric decoder. For smaller rotation angles, the infidelity from $\tilde{\mathcal{G}}_{U}$ is slightly larger than the infidelity arising from $\mathcal{G}_{U,sym}$. The two differ by at most a factor of 7 in the small $\theta$ limit. The infidelity obtained by applying the hard decoding optimization algorithm to the unperturbed channel is lowest for all sampled values of $\theta$. In (b), it can be observed that applying the decoder chosen by our optimization algorithm for the unperturbed channel to the perturbed channel results in a lower infidelity than applying the symmetric decoder to the perturbed channel, for all sampled values of $\lambda$. This indicates that our decoding scheme is very robust to small perturbations of the amplitude-phase damping channel.}
\label{fig:7qubitPerturbedInfidelityPlots}
\end{figure*}
We applied our algorithm to $\mathcal{N}_p$ and 
$\mathcal{N}_{U,p}$, giving the effective process matrices 
$\mathcal{G}(\mathcal{N}_p)$ and $\mathcal{G}(\mathcal{N}_{U,p})$ respectively.
We then applied the symmetric decoder and the decoder optimized for 
$\mathcal{N}_p$ to the perturbed noise $\mathcal{N}_{U,p}$ to obtain the process
matrices $\mathcal{G}_{U,sym}$ and $\tilde{\mathcal{G}}_{U}$ respectively. 
The infidelities of these process matrices (at the first concatenation level)
are plotted in \cref{fig:7qubitPerturbedInfidelityPlots}~(a) for the \codepar{5,1,3}
code with coherent noise and $f(\theta) = \sin^2\theta/10$ and in 
\cref{fig:7qubitPerturbedInfidelityPlots}~(b)
for the Steane code with amplitude-phase damping, $p=0.2$ and $f(\lambda) = \lambda/10$. For
both plots we averaged the values over 100 uniformly random unitaries. 
These results demonstrate that the significant improvements obtained using the 
optimized decoder are, in most studied cases, robust to perturbations in the 
noise.

The one exception we observed is for coherent noise in the \codepar{5,1,3} code
for $\theta \lesssim 0.185$, where the infidelity obtained using the optimized
decoder for the unperturbed channel is larger than that obtained using the
symmetric decoder by a factor of at most 7.

\section{Conclusion}

In this paper, we presented an optimized hard decoding algorithm for arbitrary 
local Markovian noise and numerical techniques to characterize thresholds 
for noise models. Block-wise two-qubit correlated noise was also considered. Using the analytical tools of \cref{sec:ErrorCorrectionProtocolProcessMat}, we provide numerical results in \cref{sec:AmpPhaseDamp,sec:CoherentErrorThresholdAnalysis,sec:CorrelatedNoiseThresholdAnalysis,subsec:PauliTwirl,subsec:SensitivityDecoder} which shows substantial 
improvements obtained by our algorithms compared to a fixed decoder for a 
variety of noise models, including coherent errors, correlated dephasing and 
amplitude-phase damping, and codes, namely, 
the \codepar{5,1,3}, Steane, Shor and surface-17 codes. For coherent noise, our 
optimized decoding algorithm allowed, in some cases, the noise to be 
corrected for all sampled rotation angles and reduced infidelities at a fixed 
concatenation level by orders of magnitude.

Our hard decoding algorithm is scalable and efficiently optimizes the recovery 
operations independently at each concatenation level while taking advantage of 
a code's transversal gates. At a given concatenation level, all syndrome 
measurements are considered rather than being sampled from a distribution, so 
that the performance is exactly characterized rather than containing 
statistical (and state-dependent) uncertainties.

In contrast to hard decoding, message-passing algorithms~\cite{Poulin2006} 
can increase thresholds for Pauli noise, in some cases nearing the hashing 
bound subject to sampling 
uncertainties. Large codes can also be studied using tensor 
networks~\cite{Darmawan2016}, although this requires a tensor-network 
description of 
the code and is exponential in the code distance. An interesting and important 
open problem is to combine the current techniques with those of 
Refs.~\cite{Poulin2006,Darmawan2016} to either reduce statistical uncertainties 
in 
message-passing algorithms by exploiting symmetries in the code or to treat 
larger, non-concatenated code families.

Further, we showed that performing a Pauli twirl can increase or 
decrease the threshold depending on the code and noise properties. In 
\cite{Gutierrez2015}, the Pauli twirl was found to have little impact on the 
performance 
of amplitude damping, which is known to be ``close'' to Pauli noise (that is, 
exhibit similar worst-case errors)~\cite{Kueng2016}. We conjecture that Pauli 
twirling will generally reduce thresholds for codes that have many transversal 
gates, but may improve performance for codes with fewer transversal gates.

Lastly, we considered the robustness of our hard decoding optimization 
algorithm to noise channels that were not perfectly known. We showed that by 
optimizing our decoder for a channel that was slightly perturbed by a random 
unitary operator from the 
actual channel acting on the qubits, it was still possible to obtain improved 
error rates over the symmetric decoder. However, there are some circumstances 
where the optimized decoder, while still being robust, is outperformed by the 
symmetric decoder. Determining the robustness of decoders is an open problem 
that will be especially relevant when decoders are used for experimental 
systems with incompletely characterized noise.

In Refs.~\cite{Gutierrez2015,Gutierrez2016}, the process matrix formalism was 
used to obtain 
pseudo-thresholds for the Steane code using the standard CSS decoder. 
Measurement errors were taken into account, resulting in more accurate 
pseudo-threshold values. Our methods were developed assuming that the encoding 
and decoding operations were perfect. The next step in our work will be to generalize our results to include measurement and state-preparation errors.

\section{Acknowledgements}
C. C. would like to acknowledge the support of QEII-GSST. C. C. would also like to thank Tomas Jochym-O'Connor for useful discussions and Steve Weiss for providing the necessary computational resources. This research was 
supported by the U.S. Army Research Office through grant W911NF-14-1-0103, 
CIFAR, NSERC, and Industry~Canada.

\clearpage
\appendix

\setcounter{secnumdepth}{0}
\section{Appendix: $\alpha$ and $\beta$ coefficients in closed form}
\label{app:AalphaBeta}

In this section we provide an alternative derivation of the $\alpha$ and $\beta$ coefficients found in \cref{eq:AlphaCoefficient} and \cref{eq:BetaIndividualSyn}. The latter coefficients will be given in terms of the symplectic vector representation of Pauli operators. For the bit strings $a=(a_{1},\hdots,a_{n})$ and $b=(b_{1},\hdots,b_{n})$, we will write 
\begin{align}
Z(a)X(b) = (Z^{a_{1}}\otimes \hdots \otimes Z^{a_{n}})(X^{b_{1}}\otimes \hdots \otimes X^{b_{n}}).
\label{eq:ClosedFormNotation}
\end{align}

Since the $\alpha$ coefficient is related to the overall sign of the operator $S_{k}\overline{\tau}$ (see \cref{eq:EncodingDef} and \cref{eq:EDsigmaTensorProduct}), the goal is to obtain an expression relating the overall sign of $S_{k}\overline{\tau}$ to its symplectic vector representation. An operator $S_{k} \in \mathcal{S}$ can always be written as a product of the codes stabilizer generators so that
\begin{align}
S_{k} = g_{j_{1}}\hdots g_{j_{k}},
\label{eq:SkProductGens}
\end{align}
where
\begin{align}
g_{j_{i}} = Z(a_{j_{i}})X(b_{j_{i}}).
\label{eq:GensClosedForm}
\end{align}
Defining
\begin{align}
\overline{a} &= a_{j_{1}} + \hdots + a_{j_{k}} \pmod{2} \\
\overline{b} &= b_{j_{1}} + \hdots + b_{j_{k}} \pmod{2},
\label{eq:abar}
\end{align}
we commute all the $Z$ operators in \cref{eq:SkProductGens} to the left, allowing us to write $S_{k}$ as in \cref{eq:ClosedFormNotation} 
\begin{align}
S_{k} = (-1)^{f(a_{j_{1}},\hdots , a_{j_{k}};b_{j_{1}},\hdots ,b_{j_{k}})}Z(\overline{a})X(\overline{b}).
\label{eq:SkFinal}
\end{align}
The overall sign can be obtained from the function $f$, which is given by
\begin{align}
f(a_{j_{1}},\hdots , a_{j_{k}};b_{j_{1}},\hdots ,b_{j_{k}}) = \sum_{l=1}^{k-1} 
\sum_{t=l+1}^{k}b_{j_{l}} a_{j_{t}}.
\label{eq:fFunction}
\end{align}

Writing the logical Pauli operator $\overline{\tau}$ as 
\begin{align}
\overline{\tau} = Z(\tau_{z})X(\tau_{x}),
\label{eq:logicalTau}
\end{align}
$S_{k}\overline{\tau}$ can then be written as 
\begin{align}
S_{k}\overline{\tau} = (-1)^{f(a_{j_{1}},\hdots , a_{j_{k}};b_{j_{1}},\hdots ,b_{j_{k}}) + \overline{b} \cdot \tau_{z}}Z(\overline{a} + \tau_{z})X(\overline{b} + \tau_{x}).
\label{eq:SkTauFinal}
\end{align}

For any $\mu_{j} \in \{ I,X,Y,Z \}$, we can write $\mu_{j}$ in terms of $X$ and $Z$ Pauli operators:
\begin{align}
\mu_{j} = (-i)^{a_{j} b_{j}}Z^{a_{j}}X^{b_{j}},
\label{eq:muPauliRep}
\end{align}
allowing us to write 
\begin{align}
Z(\overline{a} + \tau_{z})X(\overline{b} + \tau_{x}) = i^{(\overline{a}+\tau_{z})\cdot (\overline{b}+\tau_{x})} \mu_{1} \otimes \hdots \otimes \mu_{n}.
\label{eq:PauliFormZX}
\end{align}
It is important to note that the dot product in the factor of $i$ is \textit{not} added modulo 2. 

Using \cref{eq:PauliFormZX}, we have 
\begin{align}
S_{k}\overline{\tau} = (-1)^{f(a_{j_{1}},\hdots , a_{j_{k}};b_{j_{1}},\hdots ,b_{j_{k}}) + \overline{b} \cdot \tau_{z}} i^{(\overline{a}+\tau_{z})\cdot (\overline{b}+\tau_{x})} \mu_{1} \otimes \hdots \otimes \mu_{n}.
\label{eq:SkTauPauliForm}
\end{align}
Since the $\alpha$ coefficient takes into account the overall sign of the product between elements in the stabilizer group and the logical operators, we have 
\begin{align}
\alpha_{\phi(S_{k}\overline{\tau})}^{\tau} = \frac{1}{2^{\frac{n}{2}-1}}(-1)^{f(a_{j_{1}},\hdots , a_{j_{k}};b_{j_{1}},\hdots ,b_{j_{k}}) + \overline{b} \cdot \tau_{z}} i^{(\overline{a}+\tau_{z})\cdot (\overline{b}+\tau_{x})},
\label{eq:AlphaCoeffAppendix}
\end{align}
where the normalization factor arises from choosing a trace orthonormal basis in the sum of $E_{\tau}$.

Given a recovery map $R_{l}$ for the syndrome measurement $l$, we can write it in terms of its symplectic vector representation as 
\begin{align}
R_{l} = Z(a_{l})X(b_{l}).
\label{eq:RecoverylSymplectic}
\end{align}
From \cref{eq:GcomponentFinal} and \cref{eq:EDsigmaTensorProduct}, the $\beta$ coefficient corresponding to the recovery map $R_{l}$ can be obtained by commuting $R_{l}$ to the left of $S_{k}\overline{\tau}$ and using $R_{l}^{\dagger} R_{l} = I$. Doing so, we find that 
\begin{align}
\beta^{\tau}_{\phi(S_{k}\overline{\tau})} (R_{l})= \alpha^{\tau}_{\phi(S_{k}\overline{\tau})} (-1)^{a_{l} \cdot (\overline{b} + \tau_{x}) + b_{l} \cdot (\overline{a} + \tau_{z})}.
\label{eq:BetaCoefficientAppendix}
\end{align}

\end{document}